\documentclass[paper=a4,abstracton,numbers=noenddot]{scrartcl}

\usepackage[english]{babel}  
\usepackage{hyperref}
\usepackage[intlimits]{amsmath}
\usepackage{amsthm}
\usepackage{amssymb} 
\usepackage{units} 
\usepackage{amsfonts,mathrsfs,bbm}
\allowdisplaybreaks[1]
\usepackage[numbers]{natbib}
\usepackage{graphicx}
\usepackage[space]{grffile} % + graphicx: allow figure names with spaces
\usepackage{tikz}
\usepackage{pgfplots}
\usepackage[T1]{fontenc}
\usepackage{mathptmx}
\usepackage[scaled=.92]{helvet}
\usepackage{courier}

\usepackage[a4paper,left=2.5cm,right=2.5cm,top=3cm,bottom=2.5cm]{geometry}
\usepackage{authblk}
\usepackage{amsmath,bm}
\usepackage{graphicx}
\usepackage{mathrsfs}
\usepackage{amssymb}
\usepackage{physics}
\usepackage{tensor}
\usepackage{xcolor}

\theoremstyle{remark}
\newtheorem*{remark}{\textbf{Remark}}

\title{Geometrically exact static isogeometric analysis of arbitrarily curved plane Bernoulli-Euler beam }

\author[1,2]{A. Borkovi\'{c}}
\author[1]{B. Marussig}
\author[2,3]{G. Radenkovi\'{c}}

\affil[1]{Institute of Applied Mechanics, Graz University of Technology, Technikerstraße 4/II, 8010 Graz, Austria, aborkovic@tugraz.at, aleksandar.borkovic@aggf.unibl.org}
\affil[2]{University of Banja Luka, Faculty of Architecture, Civil Engineering and Geodesy, Department of Mechanics and Theory of Structures, 78000 Banja Luka, Bosnia and Herzegovina}
\affil[3]{Faculty of Civil Engineering, University of Belgrade, Bulevar kralja Aleksandra 73, 11000 Belgrade, Serbia}
\date{}                     %% if you don't need date to appear
\begin{document}

	\newcommand{\ssub}[2]{{#1}_{#2}} % scalar with subscripts
	\newcommand{\vsub}[2]{\textbf{#1}_{#2}} % vector with subscripts
	\newcommand{\ssup}[2]{{#1}^{#2}} % scalar with superscripts
	\newcommand{\vsup}[2]{\textbf{#1}^{#2}} % vector with superscripts
	\newcommand{\ssupsub}[3]{{#1}^{#2}_{#3}} % scalar with subscripts and superscripts
	\newcommand{\vsupsub}[3]{\textbf{#1}^{#2}_{#3}} % vector with subscripts and superscripts
	
	\newcommand{\veq}[1]{\bar{\textbf{#1}}} % vector at equidistant surface
	\newcommand{\seq}[1]{\bar{#1}} % scalar at equidistant surface
	\newcommand{\ve}[1]{\textbf{#1}} % scalar at equidistant surface
	\newcommand{\sdef}[1]{#1^*} % scalar at deformed configuration
	\newcommand{\vdef}[1]{{\textbf{#1}}^*} % vector at deformed configuration
	\newcommand{\vdefeq}[1]{{\bar{\textbf{#1}}}^*} % vec at deformed configuration at equidistant surface
	\newcommand{\trans}[1]{\textbf{#1}^\mathsf{T}} % vec or mat transpose
	\newcommand{\transmd}[1]{\dot{\textbf{#1}}^\mathsf{T}} % vec or mat transpose + material derivative
	\newcommand{\mdvdef}[1]{\dot{\textbf{#1}}^*} % material derivative of a vector at deformed configuration
	\newcommand{\mdsdef}[1]{\dot{#1}^*} % material derivative of a scalar at deformed configuration
	\newcommand{\mdv}[1]{\dot{\textbf{#1}}} % material derivative of a vector 
	\newcommand{\mds}[1]{\dot{#1}} % material derivative of a scalar
	
	\newcommand{\loc}[1]{\hat{#1}} % local, curvilienar components
	%from beginning
	\newcommand{\md}[1]{\dot{#1}} % material derivative of a scalar at deformed configuration

	\newcommand{\ii}[3]{{#1}^{#2}_{#3}} % quantity with superscript and subscript
	\newcommand{\iv}[3]{\textbf{#1}^{#2}_{#3}} % vector with superscript and subscript
	\newcommand{\idef}[3]{{#1}^{* #2}_{#3}} % quantity with superscript and subscript at deformed configuration, i - indices
	\newcommand{\ivdef}[3]{\textbf{#1}^{* #2}_{#3}} % vector with superscript and subscript at deformed configuration, i - indices
	\newcommand{\iloc}[3]{\hat{#1}^{#2}_{#3}} % local, curvilinear components
	\newcommand{\ieq}[3]{\bar{#1}^{#2}_{#3}} % quantity at equdistant surface
	\newcommand{\ic}[3]{\tilde{#1}^{#2}_{#3}} % quantity at convective frame
	\newcommand{\icdef}[3]{\tilde{#1}^{* #2}_{#3}} % quantity at convective frame
	\newcommand{\iveq}[3]{\bar{\textbf{#1}}^{#2}_{#3}} % vector at equdistant surface
	\newcommand{\ieqdef}[3]{\bar{#1}^{* #2}_{#3}} % deformed quantity at equdistant surface
	\newcommand{\iveqdef}[3]{\bar{\textbf{#1}}^{* #2}_{#3}} % deformed vector at equdistant surface
	
	\newcommand{\ieqmddef}[3]{\dot{\bar{#1}}^{* #2}_{#3}} % material derivative of a scalar at deformed configuration at equdistant surface
	\newcommand{\icmddef}[3]{\dot{\tilde{#1}}^{* #2}_{#3}} % material derivative of a scalar at deformed configuration at equdistant surface
	\newcommand{\iveqmddef}[3]{\dot{\bar{\textbf{#1}}}^{* #2}_{#3}} % material derivative of a vector at deformed configuration at equdistant surface
	
	\newcommand{\ieqmd}[3]{\dot{\bar{#1}}^{#2}_{#3}} % material derivative of a scalar at equdistant surface
	\newcommand{\icmd}[3]{\dot{\tilde{#1}}^{#2}_{#3}} % material derivative of a scalar at equdistant surface
	\newcommand{\iveqmd}[3]{\dot{\bar{\textbf{#1}}}^{#2}_{#3}} % material derivative of a vector at equdistant surface
	
	\newcommand{\imddef}[3]{\dot{#1}^{* #2}_{#3}} % material derivative of a scalar at deformed configuration
	\newcommand{\ivmddef}[3]{\dot{\textbf{#1}}^{* #2}_{#3}} % material derivative of a vector at deformed configuration
	
	\newcommand{\imd}[3]{\dot{#1}^{#2}_{#3}} % material derivative of a scalar
	\newcommand{\ivmd}[3]{\dot{\textbf{#1}}^{#2}_{#3}} % material derivative of a vector
	
	\newcommand{\iii}[5]{^{#2}_{#3}{#1}^{#4}_{#5}} % quantity with superscript and subscript on both sides
	\newcommand{\iiv}[5]{^{#2}_{#3}{\textbf{#1}}^{#4}_{#5}} % vector with superscript and subscript on both sides
	\newcommand{\iivn}[5]{^{#2}_{#3}{\tilde{\textbf{#1}}}^{#4}_{#5}} % vector with superscript and subscript on both sides
	\newcommand{\iiieq}[5]{^{#2}_{#3}{\bar{#1}}^{#4}_{#5}} % quantity at eq surface with superscript and subscript on both sides
	\newcommand{\iiieqt}[5]{^{#2}_{#3}{\tilde{#1}}^{#4}_{#5}} % quantity at eq surface with superscript and subscript on both sides
	
	\newcommand{\eqqref}[1]{Eq.~\eqref{#1}} % referencing equations with Eq. ()
	\newcommand{\fref}[1]{Fig.~\ref{#1}} % referencing figure with Fig. ()

	\maketitle
	
\section*{Abstract}

We present a geometrically exact nonlinear analysis of elastic in-plane beams in the context of finite but small strain theory. The formulation utilizes the full beam metric and obtains the complete analytic elastic constitutive model by employing the exact relation between the reference and equidistant strains. Thus, we account for the nonlinear strain distribution over the thickness of a beam. In addition to the full analytical constitutive model, four simplified ones are presented. Their comparison provides a thorough examination of the influence of a beam's metric on the structural response. We show that the appropriate formulation depends on the curviness of a beam at all configurations. Furthermore, the nonlinear distribution of strain along the thickness of strongly curved beams must be considered to obtain a complete and accurate response.

\textbf{Keywords}: Bernoulli-Euler beam; strongly curved beams; geometrically exact analysis; analytical constitutive relation

\section{Introduction}

Curved beams are fundamental structural components and their analysis is a classic topic of mechanics \cite{1992dill}. Although some analytical solutions exist, see, e.g., \cite{2011kimiaeifar}, numerical methods are inevitable for dealing with the complex nonlinear behavior of beams. The finite element method (FEM) is the most versatile numerical procedure for solving partial differential equations. The first research on the geometrically-exact FEM analysis of curved spatial beams was presented in a seminal paper by Reissner \cite{1981reissner}. This theory has been enhanced in a series of papers \cite{1985simo, 1995ibrahimbegovic, 1999crisfield, 2001atluri, 2019meier}. In \cite{1985simo}, Simo introduced the term \emph{geometrically exact beam theory} to designate that the governing equations of motion are valid regardless of the magnitude of kinematic quantities. The papers referred to are mainly focused on the shear-deformable beam model, and it is not until recently that the Bernoulli-Euler (BE) spatial model was scrutinized \cite{2013greco, 2014meier, 2015meier}. One reason for this is that the equations of the Simo-Reissner beam require only $C^0$ interelement continuity. On the other hand, the BE model requires $C^1$ interelement continuity - a requirement not fulfilled by standard Lagrange polynomials commonly used in FEM. Herein, we resolve this issue by using isogeometric analysis (IGA), \cite{2005hughes, 2017mono}. This approach utilizes (smooth) splines as basis functions for the spatial discretization of the weak form of the equations of motion, and hence, it allows for an arbitrarily high continuity between elements. 

Several works consider nonlinear shear-deformable spatial beams in the framework of IGA, \cite{2016marino, 2017marino, 2017weeger, 2019marino, 2020vob, 2020tasora, 2020choi}. In the context of the BE theory, a torsion- and rotation-free spatial cable formulation is developed in \cite{2013raknes} with attractive nonlinear dynamic applications. A spatial BE beam modeled as a ribbon with four degrees of freedom (DOF) is introduced in \cite{2013greco}, while a consistent tangent operator is derived in \cite{2015greco}. Based on \cite{2013greco}, the authors in \cite{2016bauera} discuss nonlinear applications of a spatial curved BE beam. Planar nonlinear BE beams are analyzed in \cite{2016huang, 2020vo, 2018maurin}. Different approaches to multi-parch modeling are considered in \cite{2016huang, 2020vo}, while the paper \cite{2018maurin} focuses on the weighted residuals and the collocation of strong form.

When the beam is curved, the strain distribution over the cross section becomes nonlinear. This leads to the coupling of axial and bending actions, which is readily ignored in the mentioned references in which simple decoupled constitutive relations are utilized. Nonlinear distribution of strain is evident from \textit{the initial curvature correction term} $g_0=1-\eta K$, see \cite{2003kapania}, which relates lengths of the basis vectors at the centroid and an arbitrary point of cross section to each other. This effect becomes more prominent as the curvature of the beam axis $(K)$ and height of the cross section $(h)$ increase. The product of these two quantities $Kh$ is a parameter referred to as the \emph{curviness of beam} which changes along the beam length \cite{2018borkovicb}. In this paper, we perform a nonlinear analysis of arbitrarily curved uniform BE beams. The term \emph{arbitrarily curved} does not refer just to the variable curvature of the beam axis, but more importantly, to its curviness $Kh$. We consider geometries with large curviness such as $Kh=0.5$. In \cite{2010slivker}, curved beams are grouped into small-, medium- and big-curvature beams, depending on their curviness. A beam is said to have a small curvature if its curviness is infinitesimal, $Kh \ll 1$, while it has a medium curvature if approximately, $(Kh)^2 \ll 1$. All the others belong to the category of big-curvature, also known as \emph{strongly curved} beams. In the following, we will label a beam strongly curved when $Kh > 0.1$.

Although the basic theory of strongly curved beams has been known for a long time, see e.g. \cite{2016cazzani}, it is only recently that modern numerical techniques have been applied for the analysis of these beams. Linear analysis of plane beams is given in \cite{2016cazzani, 2018borkovicb, 2019borkovicb} within the framework of IGA, while the spatial beams are analyzed in \cite{2018radenkovicb}. To the best of our knowledge, this is the first paper that deals with the nonlinear analysis of strongly curved beams. The paper is based on our previous works \cite{2018borkovicb, 2019borkovicb, 2018radenkovicb, 2017radenkovic}. The exact metric of the plane BE beam is utilized for the derivation of the weak form of equilibrium which is solved by Newton-Raphson and arc-length methods. The strict derivation of constitutive relation allows us to derive reduced models and to compare them through numerical experiments. Comparison with existing results demonstrates that the obtained formulation is reliable for the finite rotation analysis of arbitrarily curved beams. Moreover, the approach introduces an additional level of accuracy when dealing with strongly curved beams. The present formulation is geometrically exact in the sense that it strictly defines a relation between work conjugate pairs, which allows analysis of arbitrarily large rotations and displacements \cite{1985simo}. Since both geometry and displacements are interpolated with the same spline functions, the formulation exactly describes rigid-body modes and thus it is frame invariant \cite{2012armero, 2020yang}.

The paper is structured as follows: The next section presents the fundamental relations of the beam metric. Then, the description of the BE beam kinematics is provided. The finite element formulation is given in Section 4 and numerical examples are presented in Section 5. The conclusions are delivered in the last section.

\section{Metric of the beam continuum}

A detailed and rigorous definition of the beam metric is presented in this section. Following the classical BE assumption, a cross section is rigid and remains perpendicular to the beam axis in the deformed configuration. This assumption leads to a degeneration of a 3D continuum beam model into an arbitrarily shaped line. The present analysis is performed with respect to the convective frame of reference while the complete beam kinematics is defined by the translation of the beam axis.

In the notation, lowercase and uppercase boldface letters are used for vectors and tensors or matrices, respectively. An overbar designates quantities at the equidistant line of the beam, and the asterisk sign denotes the deformed configuration. Finally, the hat symbol specifies the local component of a vector, with respect to the curvilinear coordinates. The direct and index notations are applied simultaneously, depending on the context, and the standard summation convention is adopted. Greek index letters take values of 1 and 2. Partial and covariant derivatives with respect to the convective coordinates are designated with $( )_{,\alpha}$ and $( )_{\vert \alpha}$, respectively.

The elaboration on the NURBS-based IGA modeling of curves is excluded for brevity since it is readily available in the literature. For a detailed discussion on IGA and NURBS, references \cite{2005hughes, 2009kiendl, 1995piegla} are recommended. 

\subsection{Metric of the beam axis}

\begin{figure}
	\includegraphics[width=\linewidth]{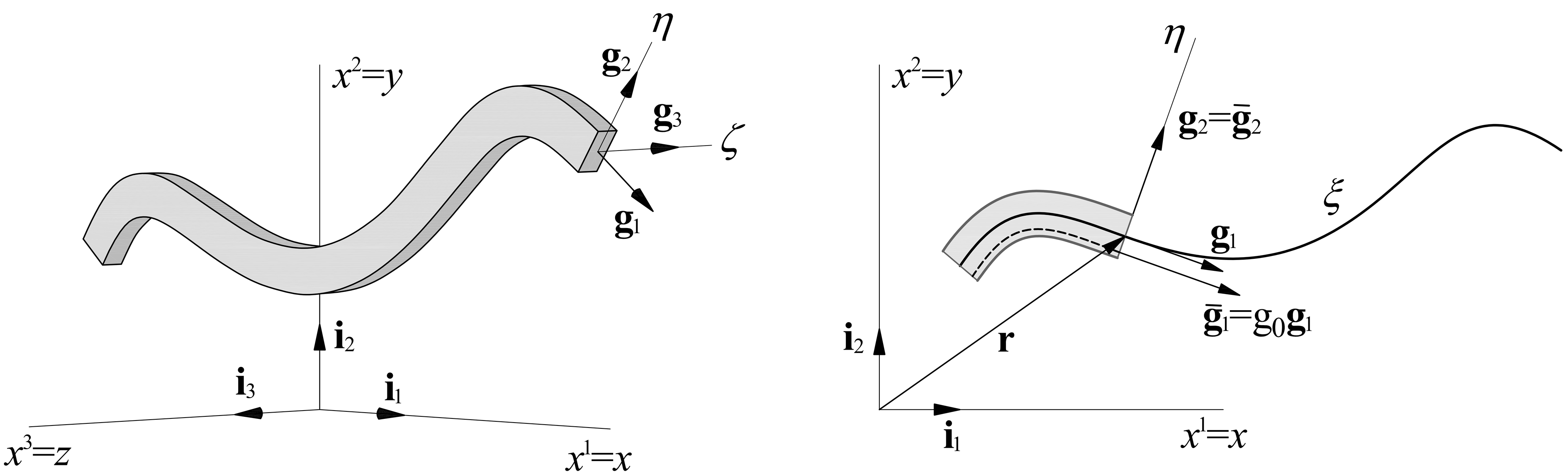}
	\caption{Degeneration of a (planar) 3D continuum into an in-plane beam. Base vectors at the centroid and at an equidistant line are depicted.}
	\label{fig:Figure 1}
\end{figure}

Let us revise some basic expressions of the metric of beam axis. A beam axis is a curve that passes through the centroids of all cross sections. It can be defined with either the arc-length coordinate $s$ or some parametric coordinate $\xi$. The position vector of the beam axis in Cartesian coordinates is $\textbf{r}=\{x=x^1, y=x^2\}$ and it is here defined as a linear combination: 
\begin{equation}
\label{eq:def:r}
\textbf{r} = x^\alpha \textbf{i}_\alpha=\sum\limits_{I=1}^{N} R_{I} (\xi) \textbf{r}_{I},  \quad   
x^\alpha = \sum\limits_{I = 1}^{N} R_{I} (\xi) x^\alpha_{I},
\end{equation}
where $R_{I}$ are univariate NURBS basis functions, $\textbf{r}_{I} = \{x_{I}, y_{I}\}$ are the position vectors of the control points $I$, and $N$ is the total number of control points \cite{2005hughes}. Furthermore, $\textbf{i}^\alpha = \textbf{i}_\alpha$ are the base vectors of the Cartesian coordinate system, see \fref{fig:Figure 1}. 
The tangent base vector of a beam axis is:
\begin{equation}
\iv{g}{}{1}=\iv{r}{}{,1}=\ii{x}{\alpha}{,1} \iv{i}{}{\alpha}.
\end{equation}
The other base vector, $\iv{g}{}{2}$, is perpendicular to $\iv{g}{}{1}$ and has unit length. Hence, we can choose the normal of the beam axis, as in \cite{2018borkovicb}, or we can rotate $\iv{g}{}{1}$ by 90 degrees and scale it. The latter approach is utilized in present formulation and the base vector $\iv{g}{}{1}$ is rotated anti-clockwise:
\begin{equation}
\label{eq: def: g2}
\iv{g}{}{2}=\frac{1}{\sqrt{g}}\Lambda \iv{g}{}{1}=-\frac{\ii{x}{2}{,1}}{\sqrt{g}} \iv{i}{}{1} + \frac{\ii{x}{1}{,1}}{\sqrt{g}} \iv{i}{}{2} = \ii{x}{\alpha}{,2} \iv{i}{}{\alpha}, \quad 
\Lambda =
\begin{bmatrix}
0 & -1\\
1 & 0
\end{bmatrix}.
\end{equation}
Here, $g$ is a component of the metric tensor and also its determinant:
\begin{align}
\label{eq:def:metric tensor axis}
\ii{g}{}{\alpha\beta}=
\begin{bmatrix}
\ii{g}{}{11} & 0\\
0 & 1
\end{bmatrix}, && \ii{g}{}{11} =\iv{g}{}{1} \cdot \iv{g}{}{1} =\det(\ii{g}{}{\alpha\beta}) =   \ii{g}{}{}.
\end{align}
This quantity relates differentials of arc-length and parametric coordinate as $\dd{s}=\sqrt{g}\dd{\xi}$. Therefore, it equals the square of Jacobian of the coordinate transformation from convective to arc-length coordinate. 

The metric of the beam axis is completely defined by the introduction of the Christoffel symbols. They follow from the differentiation of base vectors with respect to the curvilinear coordinates \cite{1973naghdi}:
\begin{equation}
\label{eq:def:christ}
\textbf{g}_{\alpha,\beta} = x^\gamma_{,\alpha\beta} \textbf{i}_\gamma =  \Gamma^{\gamma}_{\alpha \beta} \textbf{g}_\gamma 
\Rightarrow 
\Gamma^\gamma_{\alpha \beta} = x^\delta_{,\alpha\beta} x^{,\gamma}_\delta,
\end{equation}
where $\Gamma^\gamma_{\alpha \beta}$ are the Christoffel symbols of the second kind. The reciprocal base vectors of the beam axis are:
\begin{align}
\label{eq:def:reciprocal}
\iv{g}{1}{}=\ii{x}{\alpha,1}{} \iv{i}{}{\alpha}=\frac{1}{g} \iv{g}{}{1} && \textnormal{and} &&
\iv{g}{2}{}=\ii{x}{\alpha,2}{} \iv{i}{}{\alpha}=\iv{g}{}{2},
\end{align}
and the reciprocal metric tensor is:
\begin{align}
\label{eq:def:rec metric tensor axis}
\ii{g}{\alpha\beta}{}=
\begin{bmatrix}
\ii{g}{11}{} & 0\\
0 & 1
\end{bmatrix}, && \ii{g}{11}{} =\iv{g}{1}{} \cdot \iv{g}{1}{} =\det(\ii{g}{\alpha\beta}{}) = \frac{1}{g}.
\end{align}
The expression for the derivatives of base vectors in matrix form is:
\begin{equation}
\label{eq: def: derivatives of base vectors}
\begin{bmatrix}
\iv{g}{}{1,1}\\
\iv{g}{}{2,1}
\end{bmatrix}
=
\begin{bmatrix}
\iv{$\Gamma$}{1}{11} & \iv{$\Gamma$}{2}{11}\\
\iv{$\Gamma$}{1}{21} & \iv{$\Gamma$}{2}{21}
\end{bmatrix}
\begin{bmatrix}
\iv{g}{}{1}\\
\iv{g}{}{2}
\end{bmatrix}
=
\begin{bmatrix}
\iv{$\Gamma$}{1}{11} & \ic{K}{}{}\\
-K & 0
\end{bmatrix}
\begin{bmatrix}
\iv{g}{}{1}\\
\iv{g}{}{2}
\end{bmatrix},
\end{equation}
where $K$ is so-called \textit{signed curvature}:
\begin{equation}
\label{eq: def: signed curvature}
K=-\ii{\Gamma}{1}{21}=-\iv{g}{}{2,1} \cdot \iv{g}{1}{} = \frac{\ii{x}{1}{,1} \ii{x}{2}{,11}-\ii{x}{2}{,1} \ii{x}{1}{,11}}{\ii{g}{3/2}{}},
\end{equation}
and $\ic{K}{}{} = gK$ is the signed curvature of beam axis with respect to the convective coordinate frame. 

It should be noted that, since we are dealing with both the signed curvature and the modulus of curvature, the curviness $Kh$ itself can be defined with either of them. In the remainder of the paper, $Kh$ usually refers to the curviness obtained with the modulus of curvature. It is only in Subsection 5.3 that the so-called \textit{signed curviness} is used, merely for graphical representation.

\subsection{Metric of an equidistant line}

The degeneration from a 3D to a 2D beam model is trivial since we are dealing with plane beams, and all quantities are constant along the $\zeta$ coordinate, as illustrated in \fref{fig:Figure 1}. In order to reduce the beam continuum from 2D to 1D, the metric of the complete beam must be defined by some reference quantities. It is common to use the metric of the beam axis as the reference. Therefore, let us define an \emph{equidistant line} which is a set of points with $\eta=const$. Its position and tangent base vectors are:
\begin{align}
\label{eq:def:r_eq}
\begin{aligned}
\veq{r} &= \ve{r} + \eta \ve{g}_2, \\
\veq{g}_1 &= \iveq{r}{}{,1} = \ve{g}_1 - \eta K \iv{g}{}{1} = g_0 \iv{g}{}{1}, \quad \textnormal{with} \quad g_0=1-\eta K.
\end{aligned}
\end{align}
Evidently, the base vector of this line is parallel to the base vector of beam axis. Their lengths differ for the initial curvature correction term $g_0$ \cite{2003kapania}. The other base vector of the equidistant line is the same as the one of beam axis $\iveq{g}{}{2}=\iv{g}{}{2}$. The metric tensor of an equidistant line is:
\begin{equation}
\label{eq:def:gg_eq}
\ieq{g}{}{\alpha\beta}=
\begin{bmatrix}
\ii{g}{2}{0} g & 0\\
0 & 1
\end{bmatrix}, \quad \det(\ieq{g}{}{\alpha\beta}) = \ii{g}{2}{0} g = \ieq{g}{}{}.
\end{equation}
Note that the initial curvature correction term $g_0$ becomes zero for $\eta K = 1$ which is physically prohibited. Since $\eta \in \left[-h/2,h/2\right]$, it follows that the upper bound of curviness is 2.

\section{Bernoulli-Euler beam theory}

After the metric of the beam continuum is defined, the next step is to introduce a strain measure. For the convective coordinate frame, the Lagrange strain equals the difference between the current and reference metrics:
\begin{equation}
\label{eq:def:strain}
\ii{\epsilon}{}{\alpha\beta} = \frac{1}{2} \left( \idef{g}{}{\alpha \beta} - \ii{g}{}{\alpha \beta} \right).
\end{equation}
Due to the BE hypothesis, the shear strain vanishes and the position of the cross section is completely determined by the components of translation of the beam axis. In other words, these components are the only generalized coordinates of the BE beam. This fact gives rise to the so-called \emph{rotation-free} beam theories \cite{2018borkovicb}.

\subsection{Kinematics}

In the deformed configuration, the position vector of an equidistant line is:
\begin{equation}
\label{eq:def:r equidistant def}
\ieqdef{\ve{r}}{}{} = \idef{\ve{r}}{}{} (\eta) = \idef{\ve{r}}{}{} + \eta \idef{\ve{g}}{}{2},
\end{equation}
where $ \vdef{r} $ denotes the position vector of the deformed beam axis and the base vector $ \vdef{g}_2 $ is calculated similarly as in Eq.~\eqref{eq: def: g2}. The position vector of the deformed beam axis is:
\begin{equation}
\label{eq:def:r def2}
\vdef{r} = \ve{r} + \ve{u},
\end{equation}
where $ \ve{u} $ is the displacement vector of the beam axis. It is discretized with NURBS, in the same way as the geometry, \eqqref{eq:def:r}:
\begin{equation}
\label{eq:def:u}
\ve{u} = \sum\limits_{I =1}^{N} R_{I} (\xi) \ve{u}_{I}.
\end{equation}
$ \ve{u}_{I} $ is the vector of displacement components of a control point $I$ with respect to the Cartesian system. For the sake of a seamless transition to the discrete equation of motion which follows in the Section 4, let us introduce the matrix of basis functions $ \ve{N} $ such that \eqqref{eq:def:u} can be written as:
\begin{equation}
\label{eq:def:u via matrices}
\ve{u} = \ve{N} \ve{q},
\end{equation}
where:
\begin{equation}
\label{eq:def:matrices for u}
\begin{aligned}
\trans{q} &=
\begin{bmatrix}
\trans{u}_{1} & \trans{u}_{2} & ... & \trans{u}_{I} & ... & \trans{u}_{N}
\end{bmatrix}, \quad
\trans{u}_{I} = 
\begin{bmatrix}
u^1_{I} & u^2_{I}
\end{bmatrix}, 
\\
\trans{N} &= 
\begin{bmatrix}
\trans{R} & \trans{R}
\end{bmatrix}, \quad
\ve{R} = 
\begin{bmatrix}
R_{1} & R_{2} & ... & R_{I} & ... & R_{N}
\end{bmatrix}.
\end{aligned}
\end{equation}
The material derivative of the displacement field yields the velocity field, i.e.:
\begin{equation}
\label{eq:def:velocity}
\ve{v} = \mdvdef{r} = \mdv{u} = \md{u}^n \ve{i}_n = v^n \ve{i}_n = \md{x}^{*n} \ve{i}_n,
% problem with asterisk and index in superscript
\end{equation}
while its spatially discretized form follows from \eqref{eq:def:u}:  $\ve{v} = \mdv{u} = \ve{N} \mdv{q}$.
The work conjugate pair for the Cauchy stress tensor is the strain rate tensor. It is the symmetric part of the velocity gradient and equals the material derivative of \eqref{eq:def:strain}. The covariant components of the strain rate tensor, with respect to the Cartesian coordinates, can be written as:
\begin{equation}
\label{eq:def:strain rate}
	\ii{d}{}{\alpha \beta} = \imd{\epsilon}{}{\alpha \beta} = \frac{1}{2} \left( \iloc{v}{}{\alpha \vert \beta} + \iloc{v}{}{\beta \vert \alpha} \right) = \frac{1}{2} \left( \idef{x}{\delta}{,\alpha} \ii{v}{}{\delta,\beta} + \idef{x}{\delta}{,\beta} \ii{v}{}{\delta,\alpha} \right) = \frac{1}{2} \left( \ivdef{g}{}{\alpha} \cdotp \iv{v}{}{,\beta} + \ivdef{g}{}{\beta} \cdotp \iv{v}{}{,\alpha} \right),
\end{equation}
where $ \iloc{v}{}{\alpha} $ are the components of the velocity with respect to the local curvilinear coordinates \cite{2021radenkovicb}. For BE beams, the only non-zero component of the strain rate is $\ii{d}{}{11}$:
\begin{equation}
\label{eq:def: strain rate d11}
\ii{d}{}{11} = \idef{x}{\alpha}{,1} \ii{v}{}{\alpha,1} = \ivdef{g}{}{1} \cdotp \iv{v}{}{,1}.
\end{equation}

In order to define the kinematics of a beam continuum, the velocity of an equidistant line must be introduced. It is the material derivative of \eqref{eq:def:r equidistant def}:
\begin{equation}
\label{eq:def:eq velocity}
\iveqmddef{r}{}{} = \iveq{v}{}{} = \iv{v}{}{} + \eta \iv{v}{}{,2},
\end{equation}
where $\iv{v}{}{,2}$ follows from the definition of shear strain:
\begin{equation}
\label{eq:def:gradient v,2}
\ii{d}{}{12} = \frac{1}{2} \left( \ivdef{g}{}{1} \cdot \iv{v}{}{,2} + \ivdef{g}{}{2} \cdot \iv{v}{}{,1} \right)=0 \quad \Rightarrow \quad \iv{v}{}{,2} = -\frac{1}{\idef{g}{}{}} \left( \ivdef{g}{}{2} \cdot \iv{v}{}{,1}\right) \ivdef{g}{}{1} = -\frac{1}{\idef{g}{}{}} \left( \ivdef{g}{}{1} \otimes \ivdef{g}{}{2} \right) \iv{v}{}{,1}.
\end{equation}
For the purpose of further derivation, let us express the components of this gradient by index notation, \cite{2018borkovicb}:
\begin{equation}
\label{eq:def:gradient v,2 index}
\ii{v}{}{\alpha,2} = -\idef{B}{\beta}{\alpha}\ii{v}{}{\beta,1}, \quad \textnormal{with} \quad    \idef{B}{\beta}{\alpha}=\frac{1}{\idef{g}{}{}} \idef{x}{\beta}{,2} \idef{x}{}{\alpha,1} .
\end{equation}

\subsection{Strain rate at an equidistant line}

The strain rate at an equidistant line is, analogously to \eqqref{eq:def: strain rate d11}:
\begin{equation}
\label{eq:def:eq strain}
\ieq{d}{}{11} = \iveqdef{g}{}{1} \cdot \iveq{v}{}{,1}.
\end{equation}
For the evaluation of this expression, let us first define the material derivative of mixed velocity gradient, using \eqqref{eq: def: derivatives of base vectors}:
\begin{equation}
\label{eq:def: gradient v,2}
\iv{v}{}{,21} = \ivmddef{g}{}{2,1} = -\imddef{K}{}{} \ivdef{g}{}{1} -\idef{K}{}{} \iv{v}{}{,1}.
\end{equation}
With this relation and Eqs.~\eqref{eq:def:r_eq}, \eqref{eq:def:eq velocity}  and \eqref{eq:def:eq strain}, the equidistant strain rate reduces to:
\begin{equation}
\label{eq:med:eq strain rate}
\begin{aligned}
\ieq{d}{}{11} &= \left( \idef{g}{}{0} \ivdef{g}{}{1} \right) \cdot \left( \iv{v}{}{,1} + \eta \iv{v}{}{,21} \right) = \left( \idef{g}{}{0} \ivdef{g}{}{1} \right) \cdot \left( \iv{v}{}{,1} - \eta \imddef{K}{}{} \iveqdef{g}{}{1} - \eta \idef{K}{}{} \iv{v}{}{,1}  \right) \\
&=
\idef{g}{}{0} \left( \idef{g}{}{0} \ii{d}{}{11} - \eta \imddef{K}{}{} \idef{g}{}{}  \right).
\end{aligned}
\end{equation}
At this point, we will introduce a curvature change of beam axis with respect to the convective coordinate:
\begin{equation}
\label{eq:med:curvature change}
\kappa = \icdef{K}{}{} - \ic{K}{}{}  = \idef{g}{}{} \idef{K}{}{} - gK,
\end{equation}
and its material derivative, the rate of curvature change:
\begin{equation}
\imd{\kappa}{}{} = \imddef{g}{}{} \idef{K}{}{} + \idef{g}{}{} \imddef{K}{}{} = 2 \ii{d}{}{11} \idef{K}{}{} + \idef{g}{}{} \imddef{K}{}{} \quad \Rightarrow \quad \idef{g}{}{} \imddef{K}{}{} = \imd{\kappa}{}{} - 2 \ii{d}{}{11} \idef{K}{}{}.
\end{equation}
By inserting the last expression into \eqqref{eq:med:eq strain rate}, the final form of the equidistant strain rate is obtained:
\begin{equation}
\label{eq:final form of eq strain}
\ieq{d}{}{11} =
\idef{g}{}{0} \left( \idef{g}{}{0} \ii{d}{}{11} - \eta \imd{\kappa}{}{} + 2 \eta \ii{d}{}{11} \idef{K}{}{} \right) = \idef{g}{}{0} \left[ \left( 1+\eta \idef{K}{}{} \right) \ii{d}{}{11} - \eta \imd{\kappa}{}{} \right].
\end{equation}
This expression is analogous to the one derived in \cite{2018borkovicb} for the equidistant axial strain. It is valid for finite, but small, strain analysis, which falls into the scope of the geometrical exact beam theory \cite{1999crisfield}.

Finally, representing the rate of curvature change as a function of velocity gradients of the beam axis is a requirement. It follows from Eqs.~\eqref{eq: def: derivatives of base vectors}, \eqref{eq:def:gradient v,2}, and \eqref{eq:med:curvature change}:
\begin{equation}
\label{eq: def: rate of cur change}
\imd{\kappa}{}{} = \icmddef{K}{}{} = \imddef{\Gamma}{2}{11} = \ivdef{g}{}{1,1} \cdot \iv{v}{}{,2} + \ivdef{g}{}{2} \cdot \iv{v}{}{,11} = \ivdef{g}{}{2} \cdot \left( \iv{v}{}{,11} - \idef{\Gamma}{1}{11} \iv{v}{}{,1}\right).
\end{equation}

\section{Finite element formulation}

In line with the previous derivation, we will formulate the isogeometric BE element using the principle of virtual power. The generalized coordinates are the components of the velocities of the control points.

We start from the generalized Hooke law for linear elastic material, also known as the Saint Venant-Kirchhoff material model. This material model is well-suited for the small strain and large rotation analysis \cite{2004bischoff}. The only non-zero components of the stress and strain rates are related as:
\begin{equation}
\label{eq:stress strain relation}
\ieqmd{\sigma}{11}{} = E \left( \ieq{g}{11}{} \right)^2 \ieqmd{d}{}{11},
\end{equation}
where $E$ is the Young's modulus of elasticity. It is a well-known fact that this axial state of stress is an erroneous consequence of the BE assumptions. Nevertheless, it readily used due to its simplicity.

\subsection{Principle of virtual power}

The principle of virtual power represents a weak form of the equilibrium. It states that at any instance of time, the total power of the external, internal and inertial forces is zero for any admissible virtual state of motion. If the inertial effects are neglected, and body and surface loads are reduced to the beam axis, it can be written for plane BE beams as: 
\begin{equation}
\label{eq:virtual power}
\delta P = \int_{V}^{} \ieq{\sigma}{11}{} \delta \ieq{d}{}{11} \dd{V} - \int_{\xi}^{} \iloc{p}{\alpha}{} \delta \iloc{v}{}{\alpha} \sqrt{g} \dd{\xi} = \int_{V}^{} \bm{\sigma} : \delta \ve{d} \dd{V} - \int_{\xi}^{} \ve{p} \cdot \delta \ve{v} \sqrt{g} \dd{\xi} = 0,
\end{equation}
where $\bm{\sigma} $ is the Cauchy stress tensor, $\ve{d}$ is the strain rate tensor, and $\ve{p}$ is the vector of external line loads. All these quantities are measured with respect to the current, unknown, configuration. For problems with deformation-independent loads, the only requirement is to linearize the Cauchy stress. At the current $(n+1)$ configuration this stress is approximated by:
\begin{equation}
\label{eq:linearization of stress}
\iii{\bm{\sigma}}{(n+1)}{}{}{} \approx \iii{\bm{\sigma}}{(n)}{(n+1)}{}{} + \frac{{\operatorname{d}}^t \: \iii{\bm{\sigma}}{(n)}{}{}{}} {\operatorname{d}t} \Delta \ii{t}{}{(n+1)}, 
\end{equation}
where $\iii{\bm{\sigma}}{(n)}{(n+1)}{}{}$ is the stress from the previous $(n)$ configuration expressed with respect to the metric of the current $(n+1)$ configuration. 
\begin{remark}
	The additive decomposition of stress, as in \eqqref{eq:linearization of stress}, is only valid if all of the terms are expressed with respect to the same metric. Therefore, the stress from the previous configuration is here adjusted to the current metric by an \emph{adjustment term}. This term is approximately equal to the ratio of determinants of the metric tensors in two configurations, c.f.~Appendix A. The exact integration of the adjusted stress is impractical because the adjustment term consists of the ratio of the initial curvature correction terms at two configurations. Three approaches are considered here for dealing with this issue. These are discussed in detail in Appendix A and compared in Subsection 5.1.4.
\end{remark}

An appropriate time derivative in \eqqref{eq:linearization of stress} is designated with $\operatorname{d}^t () / \operatorname{d}t$ while $\Delta \ii{t}{}{(n+1)} = \ii{t}{}{(n+1)} - \ii{t}{}{(n)}$ is the time increment. The rate form of stress-strain relations requires an objective time derivative. Note that the material derivative is not objective, while the corotational (Jaumann) and convective time derivatives fulfill this property \cite{2008wriggers, 1984johnson}. The convective derivative of stress tensor results in the stress rate tensor. An important fact is that the components of the stress rate tensor are equal to the material derivatives of the components of the stress tensor, \cite{2021radenkovicb}. Having this in mind, and after the insertion of \eqqref{eq:linearization of stress} into \eqqref{eq:virtual power}, the linearized form of the principle of virtual power at the current configuration is obtained:
\begin{equation}
\label{eq:linearized virtual power}
\int_{V}^{} \ieqmd{\sigma}{11}{} \delta \ieq{d}{}{11} \dd{V} \Delta t + \int_{V}^{} \ieq{\sigma}{11}{} \delta \ieq{d}{}{11} \dd{V} = \int_{\xi}^{} \iloc{p}{\alpha}{} \delta \iloc{v}{}{\alpha} \sqrt{g} \dd{\xi}.
\end{equation}
In the remainder of this paper, we neglect the time indices and asterisks for the sake of readability. Note that this simplification does not introduce any notational ambiguity since (i) the stress and strain rates are instantaneous quantities, while the known stress is calculated at the previous configuration, and (ii) all integrations are performed with respect to the metric of the current configuration, in accordance with the updated Lagrangian procedure \cite{2007bathea}.

In order to reduce the dimension from 3D to 1D, it is necessary to integrate the left-hand side of \eqqref{eq:linearized virtual power} over the area of the cross section. Thus, integrals along the length of the beam axis are obtained: 
\begin{equation}
\label{eq:from 3D to 2D}
\begin{aligned}
\int_{V}^{} \ieqmd{\sigma}{11}{} \delta \ieq{d}{}{11} \dd{V} \Delta t + \int_{V}^{} \ieq{\sigma}{11}{} \delta \ieq{d}{}{11} \dd{V} &= \int_{\xi}^{} \left( \icmd{N}{}{} \delta \ii{d}{}{11} + \icmd{M}{}{} \delta \imd{\kappa}{}{} \right) \sqrt{g} \dd{\xi} \Delta t \\
&+ \int_{\xi}^{} \left( \ic{N}{}{} \delta \ii{d}{}{11} + \ic{M}{}{} \delta \imd{\kappa}{}{} \right) \sqrt{g}  \dd{\xi},
\end{aligned}
\end{equation}
where $\ic{N}{}{}$ and  $\ic{M}{}{}$ are the stress resultant and the stress couple, which are energetically conjugated with the reference strain rates of the beam axis, $\ii{d}{}{11}$, and $\imd{\kappa}{}{}$, while $\icmd{N}{}{}$ and $\icmd{M}{}{}$ are their respective rates: 
\begin{equation}
\label{eq: rates of section forces}
\begin{aligned}
\icmd{N}{}{} = \int_{A} (g_0)^2  \ieqmd{\sigma}{11}{} \left( 1+\eta K\right) \dd{\eta} \dd{\zeta} \quad \textnormal{and} \quad  \icmd{M}{}{} =- \int_{A} \eta (g_0)^2  \ieqmd{\sigma}{11}{} \dd{\eta} \dd{\zeta}.
\end{aligned}
\end{equation}
These expressions are analogous to those obtained in \cite{2018borkovicb}. By introducing the vectors of generalized section forces, strain rates of the beam axis, and external line loads: 
\begin{align}
\label{eq:vectors of section forces and stran rates}
\trans{f} = 
\begin{bmatrix}
\ic{N}{}{} & \ic{M}{}{}
\end{bmatrix}, &&
\trans{e} = 
\begin{bmatrix}
\ii{d}{}{11} & \imd{\kappa}{}{} 
\end{bmatrix}, &&
\trans{p} = 
\begin{bmatrix}
\ii{p}{}{1} & \ii{p}{}{2} 
\end{bmatrix},
\end{align}	
\eqqref{eq:linearized virtual power} can be written in compact matrix form as:
\begin{equation}
\label{matrix form of linearized virtual power}
\int_{\xi}^{} \transmd{f} \delta \ve{e} \sqrt{g} \dd{\xi} \Delta t + \int_{\xi}^{} \trans{f} \delta \ve{e} \sqrt{g} \dd{\xi} = \int_{\xi}^{} \trans{p}  \delta \ve{v} \sqrt{g} \dd{\xi}.
\end{equation}
This equation is nonlinear and it will be linearized in Section 4.4.

\subsection{Relation between energetically conjugated pairs}

The geometrically exact relations \eqref{eq:from 3D to 2D} and \eqref{eq: rates of section forces} are crucial for the accurate formulation of structural beam theories. In particular, they allow a rigorous definition of energetically conjugated pairs, and guarantee that the appropriate constitutive matrix is symmetric. By the introduction of Eqs.~\eqref{eq:final form of eq strain} and \eqref{eq:stress strain relation} into Eq. \eqref{eq: rates of section forces}, we obtain the exact relation between energetically conjugated pairs of stress and strain rates. The resulting symmetric constitutive matrix $\iv{D}{}{}$ is derived in \cite{2018borkovicb}:
\begin{equation}
\label{eq: def: DA}
\ivmd{f}{}{} = \iv{D}{}{} \iv{e}{}{},  \quad
\iv{D}{}{}=
\begin{bmatrix}
A & -\ieq{I}{}{} \\
-\ieq{I}{}{} & I \\
\end{bmatrix},
\end{equation}
where:
\begin{equation}
\label{eq: def: IA}
A=\int_{A}^{} \frac{\left( 1 + \eta K \right) ^2}{g_0} \dd{\eta} \dd{\zeta} , \quad
\ieq{I}{}{}=\int_{A}^{} \frac{\eta \left( 1 + \eta K \right)}{g_0} \dd{\eta} \dd{\zeta}, \quad
I=\int_{A}^{} \frac{\eta^2 }{g_0} \dd{\eta} \dd{\zeta}.
\end{equation}
Importantly, these integrals can be analytically determined for the conventional solid cross section shapes \cite{2018borkovicb}. In the following, we refer to the exact constitutive model given in \eqqref{eq: def: DA} by $D^a$.

We introduce four reduced models to examine the various influences of the exact constitutive relation. The first and the simplest of these is designated with $D^0$. It is often employed for thin beams, e.g. \cite{2020vo, 2020tasora}, since it decouples axial and bending actions:
\begin{equation}
\label{eq: def: D0}
\iv{D}{0}{}=
\begin{bmatrix}
A_0 & 0 \\
0 & I_0 \\
\end{bmatrix}, \quad
A_0=\int_{A}^{} \dd{\eta} \dd{\zeta}, \quad
I_0=\int_{A}^{} \eta^2 \dd{\eta} \dd{\zeta}.
\end{equation}
The second reduced model, $D^1$, is based on the approximation: $g_0 \rightarrow 1$. It is readily utilized for the analysis of curved beams with small curvature \cite{2013greco, 2018radenkovicb}. The model is specified by:
\begin{equation}
\label{eq: def: D1}
\begin{aligned}
\iv{D}{1}{} &=
\begin{bmatrix}
A_1 & -\ieq{I_1}{}{} \\
-\ieq{I_1}{}{} & I_1 \\
\end{bmatrix}, \quad
A_1=\int_{A}^{} \left( 1 + \eta K \right) ^2 \dd{\eta} \dd{\zeta} \approx \int_{A}^{}  \dd{\eta} \dd{\zeta}=A_0, \\
\ieq{I_1}{}{} &=\int_{A}^{} \eta \left( 1 + \eta K \right) \dd{\eta} \dd{\zeta}=KI_0, \quad
I_1=\int_{A}^{} \eta^2 \dd{\eta} \dd{\zeta}=I_0.
\end{aligned}
\end{equation}
where the quadratic term in the integrand of property $A_1$  is disregarded.

Finally, the models $D^2$ and $D^3$ are based on the Taylor approximation of the exact expressions \eqref{eq: def: IA}:
\begin{equation}
\label{eq: def: D2}
\begin{aligned}
\iv{D}{2}{} &=
\begin{bmatrix}
A_2 & -\ieq{I_2}{}{} \\
-\ieq{I_2}{}{} & I_2 \\
\end{bmatrix}, \\
A_2 &= \int_{A}^{} \dd{\eta} \dd{\zeta}= A_0, \\
\ieq{I_2}{}{} &= \int_{A}^{} \left(2 K \eta^2 \right) \dd{\eta} \dd{\zeta}=2 K I_0, \\
I_2 &= \int_{A}^{} \eta^2 \dd{\eta} \dd{\zeta}=I_0,
\end{aligned}
\end{equation}

\begin{equation}
\label{eq: def: D3}
\begin{aligned}
\iv{D}{3}{} &=
\begin{bmatrix}
A_3 & -\ieq{I_3}{}{} \\
-\ieq{I_3}{}{} & I_3 \\
\end{bmatrix}, \\
A_3 &= \int_{A}^{} \left( 1 + 4 \eta^2 K^2 \right) \dd{\eta} \dd{\zeta}= A_0 + 4 K^2 I_0, \\
\ieq{I_3}{}{} &= \int_{A}^{} \left(2 K \eta^2 + 2 \eta^4 K^3 \right) \dd{\eta} \dd{\zeta}=2 K I_0 + 2 K^3 \int_{A}^{}\eta^4 \dd{\eta} \dd{\zeta}, \\
I_3 &= \int_{A}^{} \left( \eta^2 + \eta^4 K^2 \right) \dd{\eta} \dd{\zeta}=I_0 + K^2 \int_{A}^{}\eta^4 \dd{\eta} \dd{\zeta}.
\end{aligned}
\end{equation}
where the $4^{th}$ moment of area can be easily calculated for rectangular and circular cross sections.

\subsection{Variation of strains}

Since the strain rate is a function of the generalized coordinates as well as the metric, we must vary it with respect to both arguments. By noting that the variation of the tangent vectors can be expressed as:
\begin{equation}
\label{variation of base vector g_alpha}
\delta \iv{g}{}{\alpha} = \delta \iv{v}{}{,\alpha} \Delta t,
\end{equation}
the variations of reference strains are given by:
\begin{equation}
\label{eq: variations of reference strains}
\begin{aligned}
\delta \ii{d}{}{11} &= \delta \left( \iv{g}{}{1} \cdot \iv{v}{}{,1} \right) = \delta \iv{v}{}{,1} \cdot \iv{v}{}{,1} \Delta t + \iv{g}{}{1} \cdot \delta \iv{v}{}{,1}, \\
\delta \imd{\kappa}{}{} &= \delta \left[ \iv{g}{}{2} \cdot \left( \iv{v}{}{,11} - \ii{\Gamma}{1}{11} \iv{v}{}{,1}\right) \right] \\
&= \delta \iv{v}{}{,2} \cdotp \left( \iv{v}{}{,11} - \ii{\Gamma}{1}{11} \iv{v}{}{,1}\right) \Delta t - \delta \ii{\Gamma}{1}{11} \left(\iv{g}{}{2} \cdotp \iv{v}{}{,1}\right) + \iv{g}{}{2} \cdotp \left( \delta \iv{v}{}{,11} - \ii{\Gamma}{1}{11} \delta \iv{v}{}{,1}\right).
\end{aligned}
\end{equation}
Most of the terms in the previous expression are easily computed since the variation is explicitly performed with respect to the unknown variables. However, the first two addends of $\imd{\kappa}{}{}$ are exceptions since they require the variations of the velocity of the normal and the Christoffel symbol. These variations must be represented via the variations of generalized coordinates \cite{2017radenkovic}. 

The variation of the velocity of the normal follows directly from \eqqref{eq:def:gradient v,2}:
\begin{equation}
\label{eq: variation of normal}
\delta \iv{v}{}{,2} = -\frac{1}{\ii{g}{}{}} \left( \iv{g}{}{2} \cdot \delta \iv{v}{}{,1}\right) \iv{g}{}{1}.
\end{equation}
For the variation of the Christoffel symbols, we start from \eqqref{eq:def:christ} and, after some straightforward calculation, obtain:
\begin{equation}
\label{eq:variation of chris}
\delta \ii{\Gamma}{1}{11} = \delta \left(  \iv{g}{}{1,1} \cdot \iv{g}{1}{} \right) = \left[ \frac{1}{g} \iv{g}{}{1}  \delta \iv{v}{}{,11} + \frac{1}{g} \left( \ic{K}{}{} \iv{g}{}{2} - \ii{\Gamma}{1}{11} \iv{g}{}{1} \right) \cdot \delta \iv{v}{}{,1} \right] \Delta t.
\end{equation}
By inserting Eqs.~\eqref{eq: variation of normal} and \eqref{eq:variation of chris} into \eqqref{eq: variations of reference strains}, the final expression for the variation of the rate of curvature change is found: 
\begin{equation}
\label{eq: variation of curvature change}
\begin{aligned}
\delta \imd{\kappa}{}{} &= \iv{g}{}{2} \cdotp \left( \delta \iv{v}{}{,11} - \ii{\Gamma}{1}{11} \delta \iv{v}{}{,1}\right) \\
&- \frac{1}{\ii{g}{}{}} \left\{ \left( \iv{g}{}{2} \cdot \delta \iv{v}{}{,1}\right) \iv{g}{}{1} \cdotp \left( \iv{v}{}{,11} - \ii{\Gamma}{1}{11} \iv{v}{}{,1}\right) + \left[ \iv{g}{}{1}  \delta \iv{v}{}{,11} + \left( \ic{K}{}{} \iv{g}{}{2} - \ii{\Gamma}{1}{11} \iv{g}{}{1} \right) \cdot \delta \iv{v}{}{,1} \right] \left( \iv{g}{}{2} \cdotp \iv{v}{}{,1} \right) \right\} \Delta t.
\end{aligned}
\end{equation}

\subsection{Discrete equation of motion}

In this subsection, the virtual power is spatially discretized and linearized. We start by introducing the matrix $\iv{B}{}{L}$, which relates the reference strain rates of the beam axis with the velocities of control points:
\begin{equation}
\label{eq: e=BL q}
\ve{e} = \iv{B}{}{L} \ivmd{q}{}{}.
\end{equation}
Using Eqs.~\eqref{eq:def: strain rate d11} and \eqref{eq: def: rate of cur change}, the vector of the reference strain rates  can be represented as:
\begin{equation}
\label{eq: vector of reference strains matrix form}
\ve{e} = \ve{H} \ve{w} = \sum_{\alpha=1}^{2} \iv{H}{}{\alpha} \iv{w}{\alpha}{}, \quad \ve{H} = 
\begin{bmatrix}
\iv{H}{}{1} & \iv{H}{}{2}
\end{bmatrix}, \quad
\iv{H}{}{\alpha} = 
\begin{bmatrix}
\ii{x}{}{\alpha,1} & 0 \\
-\ii{\Gamma}{1}{11} & \ii{x}{}{\alpha,2} 
\end{bmatrix},  \quad
\iv{w}{\alpha}{} = 
\begin{bmatrix}
\ii{v}{\alpha}{,1} \\
\ii{v}{\alpha}{,11} 
\end{bmatrix},
\end{equation}
where the vector $\ve{w}$ is defined by:
\begin{equation}
\label{eq:w definition}
\ve{w} = \ve{B} \ivmd{q}{}{}, \quad  \ve{w} = 
\begin{bmatrix}
\iv{w}{1}{} \\
\iv{w}{2}{} 
\end{bmatrix}, \quad \ve{B} = 
\begin{bmatrix}
\iv{B}{1}{1} & ... & \iv{B}{1}{I} & ... & \iv{B}{1}{N} \\
\iv{B}{2}{1} & ... & \iv{B}{2}{I} & ... & \iv{B}{2}{N} 
\end{bmatrix}.
\end{equation}
The submatrices $\iv{B}{\alpha}{I}$ for an arbitrary control point $I$ consist of the derivatives of the basis functions, i.e.:
\begin{equation}
\label{eq: submatrices BIJ}
\iv{B}{1}{I} = 
\begin{bmatrix}
\ii{R}{}{I,1} & 0 \\
\ii{R}{}{I,11} & 0 
\end{bmatrix}, \quad
\iv{B}{2}{I} = 
\begin{bmatrix}
0 & \ii{R}{}{I,1} \\
0 & \ii{R}{}{I,11} 
\end{bmatrix}.
\end{equation}
The matrix $\iv{B}{}{L}$ now follows as:
\begin{equation}
\label{eq: e=Hw, BL=HB}
\ve{e} = \ve{H} \ve{w} = \ve{H} \ve{B} \ivmd{q}{}{} = \iv{B}{}{L} \ivmd{q}{}{}, \quad \iv{B}{}{L} = \ve{H} \ve{B}.
\end{equation}
Next, we need to specify the explicit matrix form of the virtual power generated by the known stress and the variation of strain rate with respect to the metric, Eqs.~\eqref{matrix form of linearized virtual power} and \eqref{eq: variations of reference strains}:
\begin{equation}
\label{eq: part of vp generated by known stress and variation of strain rate}
\begin{aligned}
\int_{\xi}^{} \trans{f} \delta \iv{B}{}{L} \ivmd{q}{}{} \sqrt{g} \dd{\xi} \Delta t &= \int_{\xi}^{} \sum_{\alpha=1}^{2} \sum_{\beta=1}^{2}  \trans{$(\iv{w}{\alpha}{})$} \iv{G}{\beta}{\alpha} \delta \iv{w}{}{\beta} \sqrt{g} \dd{\xi} \Delta t \\
\sum_{\alpha=1}^{2} \sum_{\beta=1}^{2}  \trans{$(\iv{w}{\alpha}{})$} \iv{G}{\beta}{\alpha} \delta \iv{w}{}{\beta} &= \ic{N}{}{} \left( \ii{v}{\alpha}{,1} \delta \ii{v}{}{\alpha,1} \right) + \ic{M}{}{} \left( \ii{v}{\alpha}{,1} \ii{Y}{\beta}{\alpha } \delta \ii{v}{}{\beta,1} - \ii{v}{\alpha}{,11} \ii{B}{\beta}{\alpha } \delta \ii{v}{}{\beta,1} - \ii{v}{\alpha}{,1} \ic{B}{\beta}{\alpha} \delta \ii{v}{}{\beta,11} \right) ,
\end{aligned}
\end{equation}
where the matrix $\iv{G}{\beta}{\alpha}$ is: 
\begin{equation}
\label{eq:Gnm def}
\iv{G}{\beta}{\alpha} = 
\begin{bmatrix}
\ic{N}{}{} \ii{\delta}{\beta}{\alpha} + \ic{M}{}{} \ii{Y}{\beta}{\alpha} & -\ic{M}{}{} \ic{B}{\beta}{\alpha} \\
-\ic{M}{}{} \ii{B}{\beta}{\alpha} & 0 
\end{bmatrix},
\end{equation}
with:
\begin{equation}
\label{eq:Gnm elements def}
\ii{Y}{\beta}{\alpha} = \ii{\Gamma}{1}{11} \ii{B}{\beta}{\alpha} - \frac{1}{g} \left( \ic{K}{}{} \ii{x}{\beta}{,2} -\ii{\Gamma}{1}{11} \ii{x}{\beta}{,1}\right) \ii{x}{}{\alpha,2}, \quad \ic{B}{\beta}{\alpha} = \frac{1}{g} \ii{x}{\beta}{,1} \ii{x}{}{\alpha,2} = \ii{B}{\alpha}{\beta}. 
\end{equation}
By introducing the total matrix of the generalized section forces:
\begin{equation}
\label{eq: G matrix def}
\ve{G} = 
\begin{bmatrix}
\iv{G}{1}{1} & \iv{G}{2}{1} \\
\iv{G}{1}{2} & \iv{G}{2}{2} 
\end{bmatrix},
\end{equation}
expression \eqref{eq: part of vp generated by known stress and variation of strain rate} can be written as:
\begin{equation}
\label{eq:transf geometric term of VP to bilinear form}
\int_{\xi}^{} \trans{f} \delta \iv{B}{}{L} \ivmd{q}{}{} \sqrt{g} \dd{\xi} \Delta t = \int_{\xi}^{} \trans{w} \ve{G} \delta \ve{w} \sqrt{g} \dd{\xi} \Delta t.
\end{equation}
A careful inspection of Eqs.~\eqref{eq:Gnm def} and \eqref{eq: G matrix def} reveals that the matrix of generalized section forces, $\ve{G}$, is symmetric \cite{2017radenkovic}. Now, the integrands in the equation of the virtual power \eqref{matrix form of linearized virtual power} reduce to:
\begin{equation}
\label{eq: terms of VP rewritten}
\begin{aligned}
\transmd{f} \delta \ve{e} &\approx  \transmd{q} \trans{$\iv{B}{}{L}$} \ve{D} \iv{B}{}{L} \delta \ivmd{q}{}{},\\
\trans{f} \delta \ve{e}  &=  \trans{f} \left( \delta \iv{B}{}{L} \ivmd{q}{}{} + \iv{B}{}{L} \delta \ivmd{q}{}{}\right)  =  \transmd{q} \trans{B} \ve{G} \ve{B} \delta \ivmd{q}{}{} \Delta t + \trans{f} \iv{B}{}{L} \delta \ivmd{q}{}{} ,
\end{aligned}
\end{equation}
where the first term is linearized by neglecting the variation of strain rate with respect to the metric:
\begin{equation}
\label{eq: terms of VP approximation}
\delta \ve{e} \approx \iv{g}{}{1} \cdotp \delta \iv{v}{}{,1} = \iv{B}{}{L} \delta \ivmd{q}{}{}. 
\end{equation}
Finally, the equation of equilibrium reduces to:
\begin{equation}
\label{eq: virtual equilibrium}
\transmd{q} \int_{\xi}^{} \left( \trans{$\iv{B}{}{L}$} \ve{D} \iv{B}{}{L} + \trans{B} \ve{G} \ve{B} \right) \sqrt{g} \dd{\xi} \delta \ivmd{q}{}{} \Delta t = \int_{\xi}^{} \trans{p} \ve{N} \sqrt{g} \dd{\xi} \delta \ivmd{q}{}{} - \int_{\xi}^{} \trans{f} \iv{B}{}{L} \sqrt{g} \dd{\xi} \delta \ivmd{q}{}{},
\end{equation}
which is readily written in the standard form:
\begin{equation}
\label{eq:standard form of equlibrium}
\iv{K}{}{T} \Delta \ve{q} = \ve{Q} - \ve{F}, \quad \left( \Delta \ve{q} = \ivmd{q}{}{} \Delta t\right),
\end{equation}
where:
\begin{equation}
\label{eq: Kt}
\iv{K}{}{T} = \int_{\xi}^{} \trans{$\iv{B}{}{L}$} \ve{D} \iv{B}{}{L} \sqrt{g} \dd{\xi} + \int_{\xi}^{} \trans{B} \ve{G} \ve{B} \sqrt{g} \dd{\xi}, 
\end{equation}
is the tangent stiffness matrix and:
\begin{equation}
\label{eq: Q and F}
\ve{Q} = \int_{\xi}^{} \trans{N} \ve{p} \sqrt{g} \dd{\xi}, \quad \ve{F} = \int_{\xi}^{} \trans{$\iv{B}{}{L}$} \ve{f} \sqrt{g} \dd{\xi},
\end{equation}
are the vectors of the external and internal forces, respectively. The vector $\Delta \ve{q}$ in \eqqref{eq:standard form of equlibrium} contains increments of the displacements. Due to the approximation introduced in \eqqref{eq: terms of VP approximation}, the solution of \eqqref{eq:standard form of equlibrium} does not satisfy the principle of virtual power directly, but has to be enforced by additional numerical schemes. Here, both the Newton-Raphson and arc-length methods are employed.

The present derivation of the geometric stiffness matrix differs from the conventional procedure for nonlinear beam formulations \cite{2020vo}. In particular, the full BE beam metric is incorporated in \eqqref{eq: part of vp generated by known stress and variation of strain rate} and the symmetric nature of the geometric stiffness term is stressed by the elegant and compact form of Eqs.~\eqref{eq:Gnm def} and \eqref{eq:transf geometric term of VP to bilinear form}. The latter confirms that the formulation is valid and that the adopted force and strain quantities are energetically conjugated.

\section{Numerical examples}

The aim of the subsequent numerical experiments is to validate the proposed approach and to examine the influence of curviness on structural response.

The boundary conditions are imposed strongly and the rotations are treated with special care, see \cite{2018borkovicb}, since they are not utilized as DOFs. Locking issues are not considered, but the presence of membrane locking is already detected for linear analyses in \cite{2018borkovicb}, and it is present in nonlinear analysis as well. However, its influence is alleviated with the usage of higher order basis functions \cite{2014adam}. It is interesting that high interelement continuity increases the locking effect due to the presence of more constraints, which must be satisfied. This issue can be dealt with an appropriate numerical integration scheme \cite{2014adam}. The quest for optimal quadrature rules in IGA is an ongoing topic, see e.g. \cite{2012auricchioa}, but the standard Gauss quadrature with $p+1$ integration points are used here.

Since no rotational DOFs are required for BE beams, the implementation of multi-patch structures receives much attention \cite{2020voa, 2020vo, 2014greco, 2020marchiori}. A straightforward algorithm for handling multi-patch structures is implemented here \cite{2009kiendl}. The rotation at the end of NURBS curve depends on the displacements of the last two control points \cite{1995piegla}. Therefore, the rotation of two NURBS curves at a joint is a function of six displacement components. In order to prescribe rigid connection between two patches, one displacement component is constrained as a function of the other five. This approach is straightforward and does not require the introduction of bending strips or end rotational DOFs \cite{2016greco, 2020vo}. 

All the results are related to the load proportionality factor ($LPF$), rather than the load intensity itself. Although the implemented code allows the usage of large load increments for some examples, the results are mostly calculated with small increments in order to enable a clear comparison of the obtained results via smooth equilibrium paths. For problems that do not exhibit a snap behavior, it is possible to obtain the deformed configurations for the prescribed load increments. Thus, the relative differences between different models can be compared along the equilibrium path, c.f. Subsection 5.1. These graphs give a clearer insight into discrepancies of the models than a simple comparison of equilibrium paths.

Since the time is a fictitious quantity in the present static analysis, strain and stress rates are equal to strains and stresses, respectively.

\subsection{Cantilever beam subjected to an end concentrated moment}

\begin{figure}[hb]
	\includegraphics[width=8 cm]{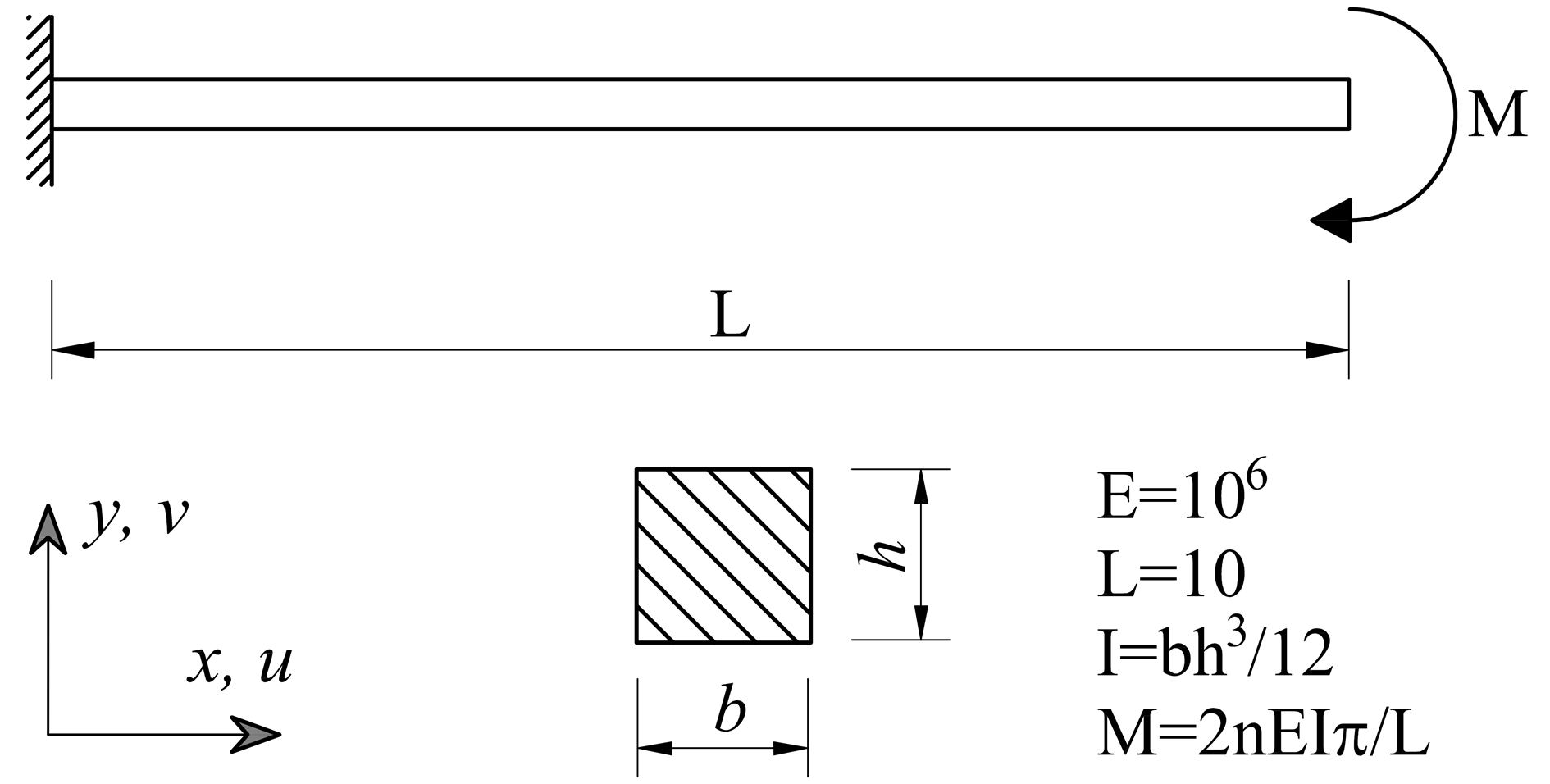}\centering
	\caption{Cantilever beam. Geometry and applied load. }
	\label{fig:mainspring: disposition}
\end{figure}

This example is a classic benchmark test for the validation of nonlinear beam formulations \cite{2020vo, 2016bauera, 2016huang}. The cantilever beam is loaded at its free end with a moment $2nEI\pi /L$, where $n$ defines the number of circles into which the beam rolls-up, Fig. \ref{fig:mainspring: disposition}. 
A square cross section is considered and its dimension varies from the set $b=h\in\{0.05, 0.1, 0.2, 0.4, 0.8\}$. If the beam is rolled up into one circle ($n=1$) the curviness becomes $Kh\approx\{0.03,0.06,0.13,0.25,0.50\}$ for $LPF=1$. If the beam is rolled up into two circles ($n=2$), the value of curviness at the final configuration is also doubled.

A moment load can be applied in various ways for rotation-free models. For example, an additional knot can be inserted to define a couple \cite{2016bauera}, or the linear stress distribution can be imposed, \cite{2020choi}. Here, no additional knots are added and the force couple is defined by the external virtual power, analogous to \cite{2018borkovicb}. Since we are dealing with rotation-free nonlinear analysis, this couple changes direction and intensity throughout the whole deformation. Therefore, the external load vector must be updated at each iteration to define the applied moment accurately. Similar to \cite{2016bauera}, the contribution from these non-conservative external forces to the tangent stiffness matrix is disregarded. 

\subsubsection{Convergence analysis}

\begin{figure}[t]
	\includegraphics[width=\linewidth]{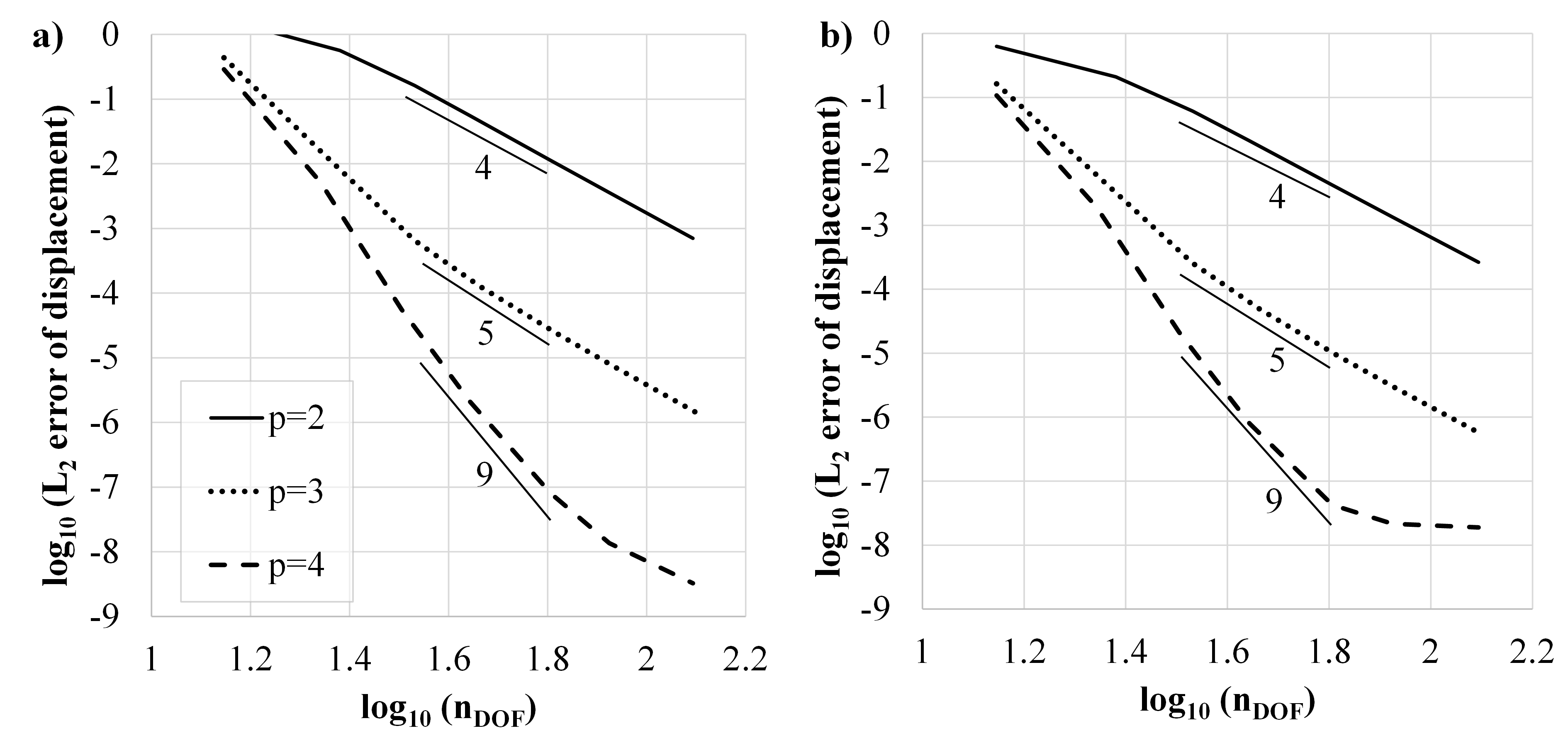}
	\caption{Cantilever beam. Convergence of displacement for LPF=1 for the models a) $D^1$, b) $D^a$. }
	\label{fig:mainspring: displacement convergence two models}
\end{figure}

We examine the convergence properties of the developed formulation for two constitutive models, i.e., $D^1$ and $D^a$, using $n=1$ and $h=0.2 \rightarrow Kh \approx 0.13$. Since these two models return different structural responses, different reference solutions must be utilized. For the $D^1$ model, reference analytical solution for thin beam is employed \cite{2016bauera}. Regarding the $D^a$ model, the analytical solution is not available in literature and a mesh of 60 quintic $C^1$ elements (484 DOFs) is adopted as a reference solution. The $L_2\textnormal{-norm}$ of the relative error of displacement for $LPF=1$ is considered in Fig. \ref{fig:mainspring: displacement convergence two models} for three different polynomial orders with highest available interelement continuity. 
Since the expected order of convergence is $p+1$, \cite{2013greco}, the obtained orders are higher. 

\begin{figure}[t]
	\includegraphics[width=\linewidth]{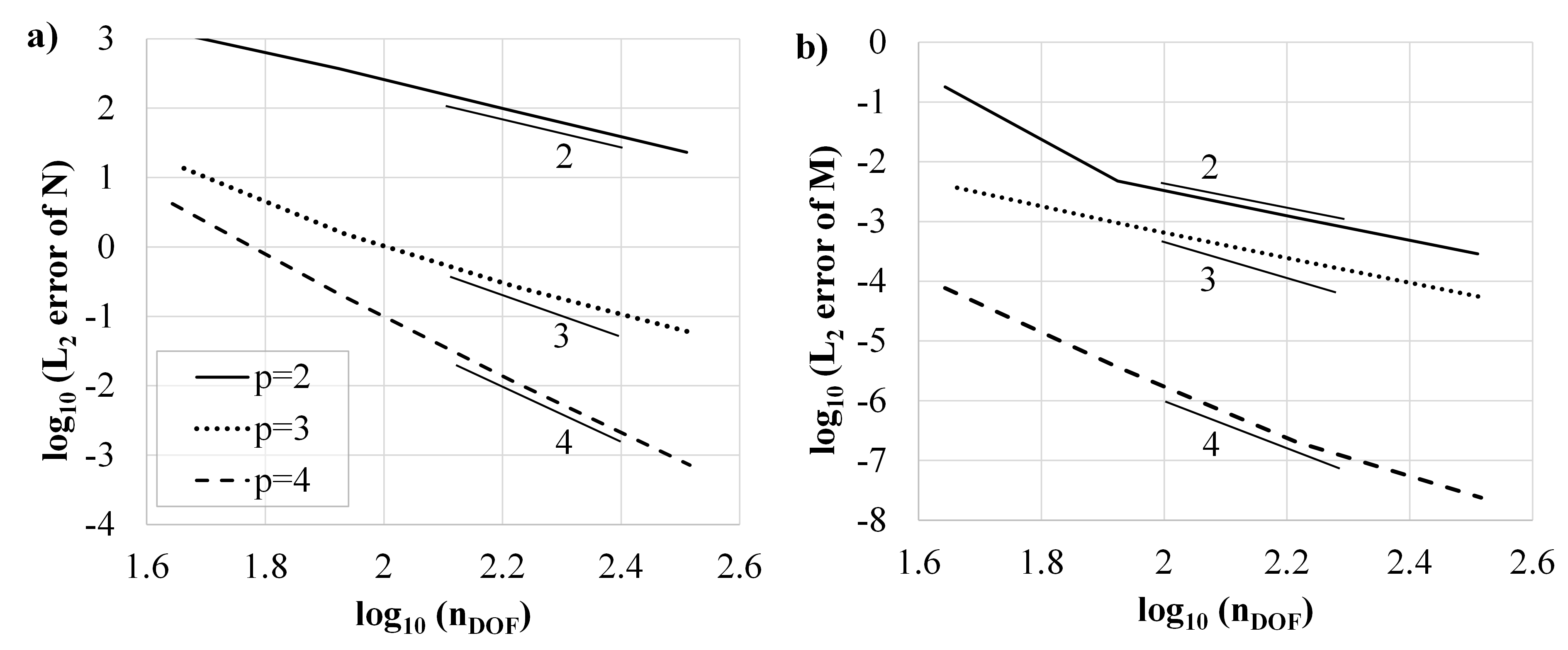}
	\caption{Cantilever beam. Convergence of: a) normal force, and b) bending moment, for LPF=1. }
	\label{fig:mainspring: force convergence two models}
\end{figure}
Fig. \ref{fig:mainspring: force convergence two models} illustrates the convergence of the normal force and the bending moment using the $D^a$ model. 
The analytical solution is used as reference. The expected order is $p$ and it is obtained in all tests. The exception is the mesh with cubic splines. A similar behavior has been observed in \cite{2016marino} and attributed to the higher order derivatives involved in the strong form. In the rest of this example, quartic splines with $C^3$ interelement continuity are exclusively used.

\subsubsection{Comparison of constitutive models}

\begin{figure}
	\includegraphics[width=\linewidth]{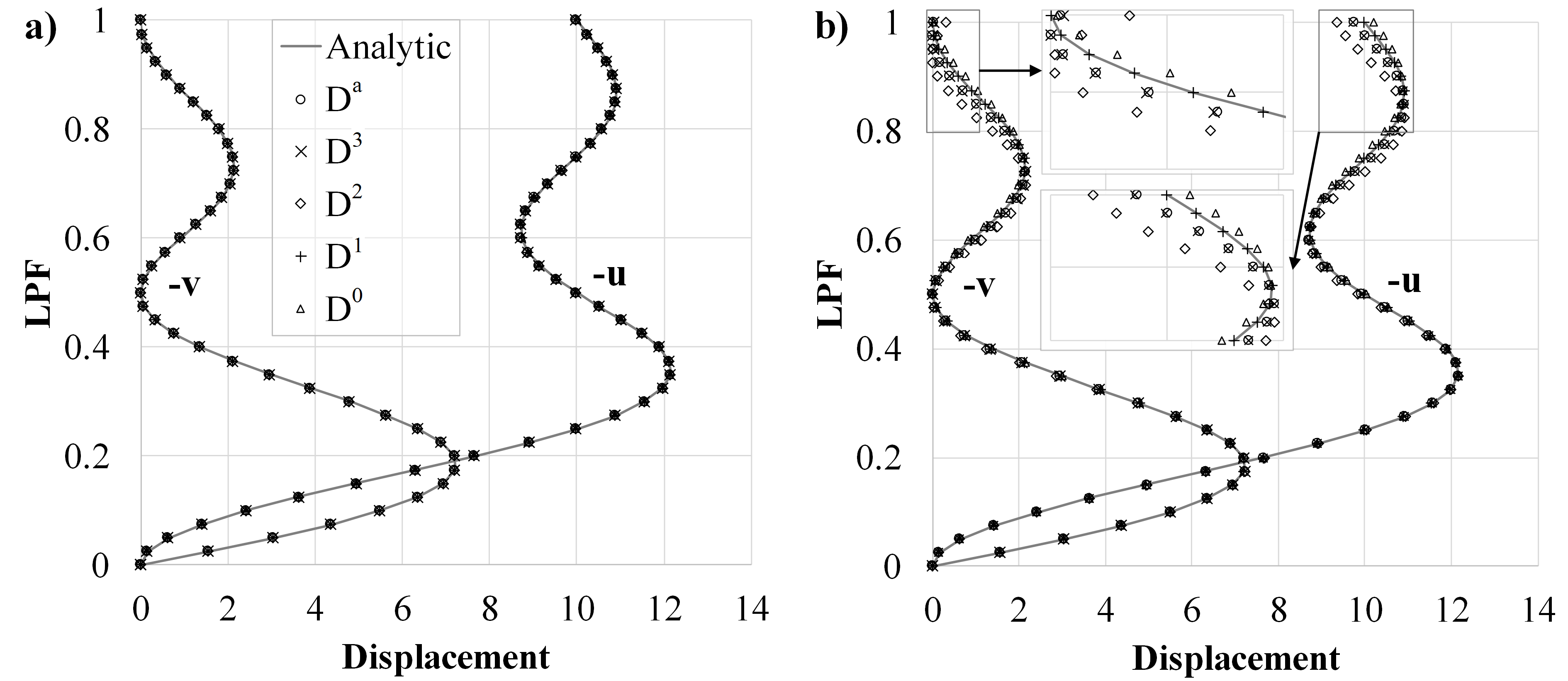}
	\caption{Cantilever beam. Comparison of the displacement components of the tip obtained by different constitutive models and analytically: a) $Kh \approx 0.06$, b) $Kh \approx 0.50$.}
	\label{fig:mainspring: displacement different models}
\end{figure}

A discretization with 24 elements is employed to calculate the equilibrium path of the tip for $n=2$ and two different values of curviness. The results are compared with the analytical ones and shown in Fig. \ref{fig:mainspring: displacement different models}. For a beam with the curviness $Kh \approx 0.06$, the results of all models are in almost full agreement with the analytical solution, Fig. \ref{fig:mainspring: displacement different models}a. 
However, when the cross section dimensions are increased such that the curviness becomes $Kh \approx 0.50$, the discrepancies become significant, Fig. \ref{fig:mainspring: displacement different models}b. The $D^1$ model is fully aligned with the analytical solution for thin beams, while the $D^0$ model slightly deviates. On the other hand, the results obtained with the $D^a$ and $D^3$ models are virtually indistinguishable, but differ from those of the thin beam. The $D^2$ model deviates from the exact predictions, but with a different sign than the $D^0$ and $D^1$ models. These results show that the present approach can return accurate results for problems with large displacements and rotations. For beams with small curvature, the results are aligned with the classic predictions. As the curviness increase, the difference between constitutive models becomes clearly visible.

\begin{figure}
	\includegraphics[width=\linewidth]{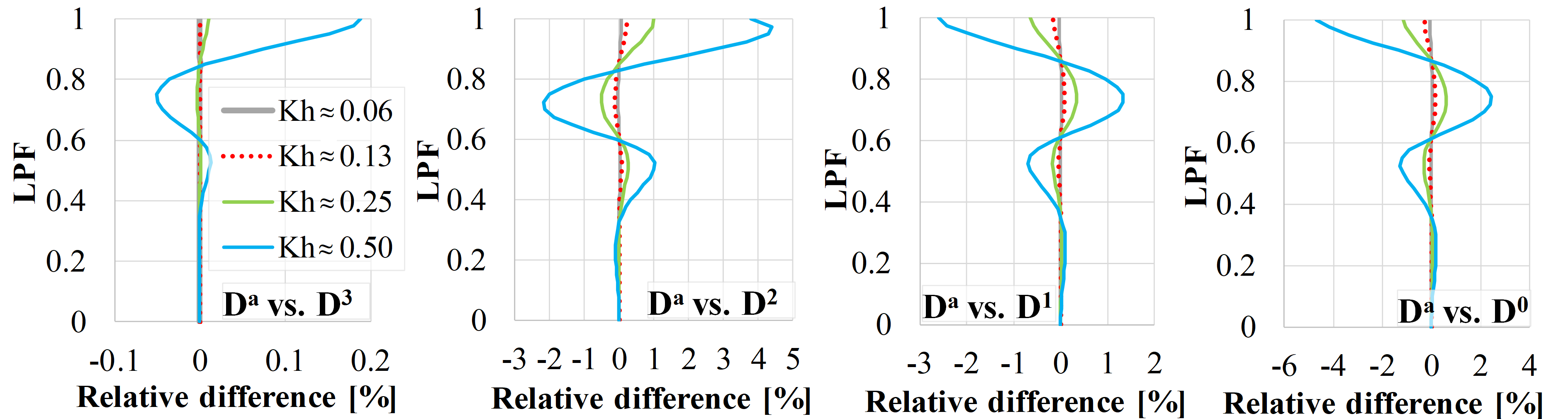}
	\caption{Cantilever beam. Relative differences of the displacement of the tip for four different values of curviness. The results by the reduced models are compared with those from the exact one. }
	\label{fig:mainspring: displacement relative difference}
\end{figure}

In order to present these findings more concisely, the relative differences of the reduced models with respect to the $D^a$ model are presented in Fig.~\ref{fig:mainspring: displacement relative difference} for four different values of curviness. 
These results confirm that the reduced model $D^3$ returns reasonably accurate results, even for strongly curved beams. It is interesting to note the relatively large differences of the $D^2$ model, which suggest that the inclusion of higher order terms, as in the reduced constitutive model $D^3$, is required if accurate results for strongly curved beams are sought, see Eqs.~\eqref{eq: def: D2} and \eqref{eq: def: D3}. 

Another important conclusion follows from this specific example. Since the curviness is constant along the beam, the introduced exact metric has more impact on the beam response than it is the case with examples where the maximum curviness is local. We can observe, e.g., that if the curviness along the whole beam is less than 0.15, the error of the simple decoupled $D^0$ model is less than 0.3 \%. As the curviness increases over 0.25, the error increase over 1.1 \%, etc.

\subsubsection{Specific aspects of strongly curved beams}
\begin{figure}
	\includegraphics[width=\linewidth]{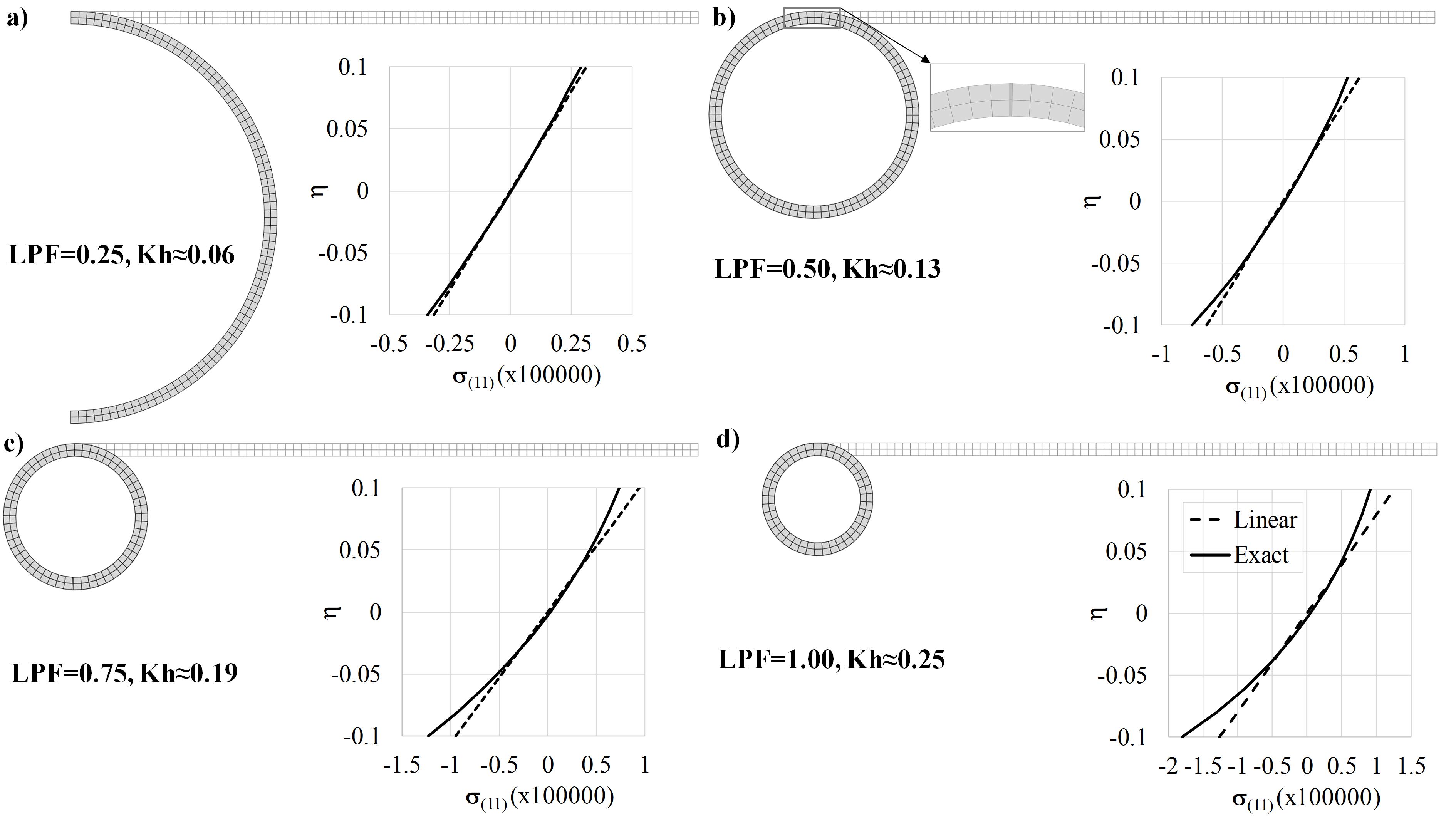}
	\caption{Cantilever beam. Reference and deformed configurations, and stress distribution across the height of cross section for four levels of load: a) $LPF=0.25$, b) $LPF=0.50$, c) $LPF=0.75$, d) $LPF=1.00$.  }
	\label{fig:mainspring: deformed configurations}
\end{figure}

This example is sometimes misinterpreted due to the pure bending conditions. It is assumed that there is no change of the length of beam axis and the beam rolls-up into a perfect circle with the circumference equal to the length of the beam \cite{2020tasora}. However, when we deal with a geometrically exact analysis that considers the effect of curviness, the strain is distributed nonlinearly across the height of the cross section with a non-zero value at the centroid. This strain is extensional and its value is negligible for very thin beams, but becomes significant for strongly curved beams. Consequently, the axis of a cantilever loaded with tip moment will reach full circle for $n<1$ and its circumference will be somewhat larger than the original length. In order to depict this, the four configurations of the beam with $h=0.2$ are visualized in Fig. \ref{fig:mainspring: deformed configurations}. 
Note how the initially thin and slender beam becomes strongly curved during the deformation. As a consequence, the metric, axial strain, and stress across the height of the cross section are distributed nonlinearly. The distribution of stress is shown next to the deformed configurations and compared with the classic linear distribution. Evidently, the intrados stress is greater, and the extrados stress is lesser than the equivalent linear stress. Due to the small extension of beam axis, the ends of the beam overlap for $LPF=0.5$ and $LPF=1$. This overlap for $LPF=0.5$ is zoomed in Fig.~\ref{fig:mainspring: deformed configurations}b.

\begin{figure}
	\includegraphics[width=\linewidth]{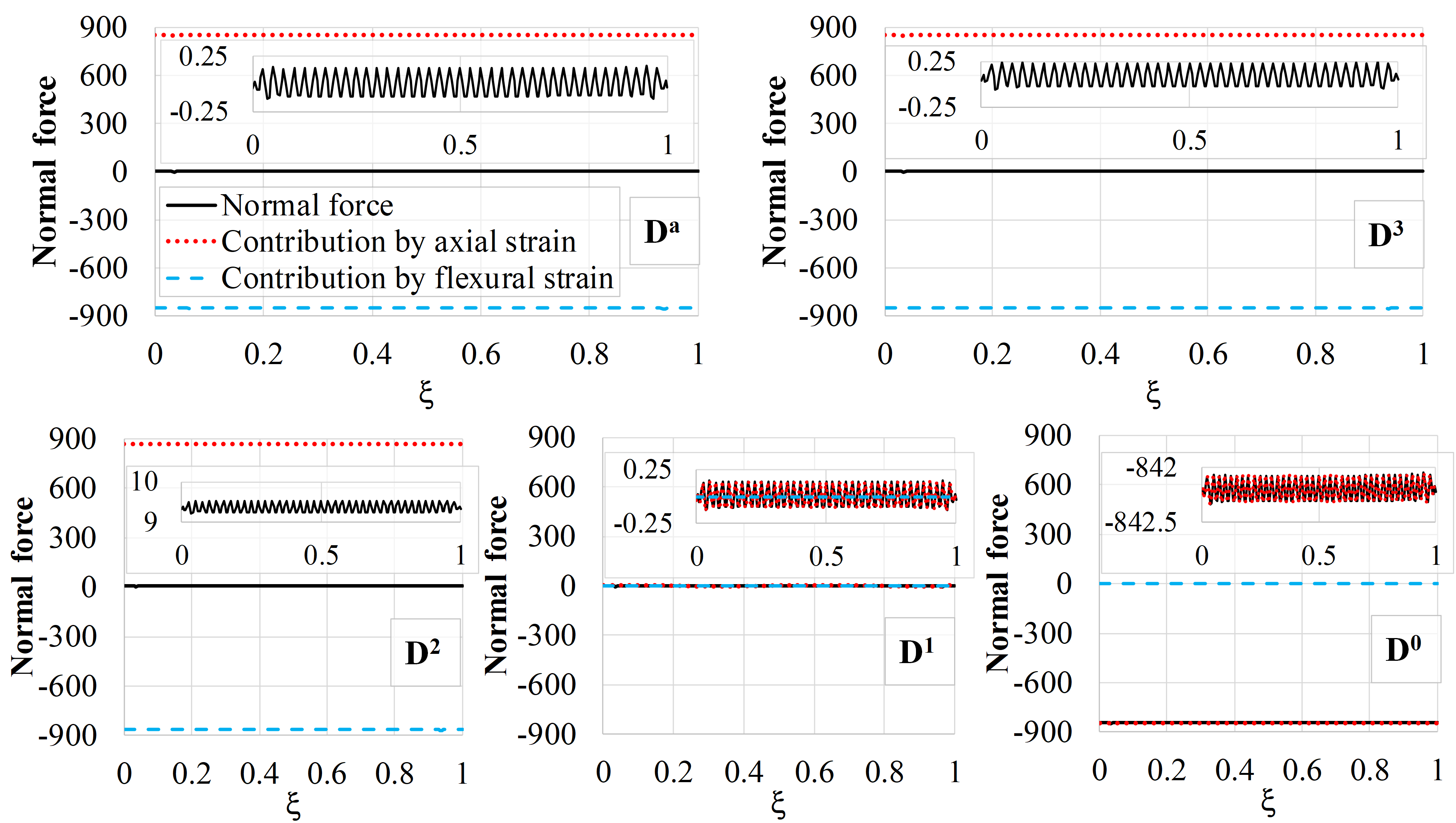}
	\caption{Cantilever beam. Comparison of normal force and its contributions by axial and flexural strain for different constitutive models. The resulting normal force is zoomed on all the graphs. }
	\label{fig:mainspring: normal force}
\end{figure}

Furthermore, due to the axial strain of the beam axis, it is necessary to include the influence of flexural strain when calculating the normal section force \cite{2018borkovicb}. To examine this effect, 40 elements are used for a model with $n=1$ and $Kh=0.25$, and the influence of both axial and flexural strains is observed. All five constitutive models are employed and the results are presented in Fig.~\ref{fig:mainspring: normal force}. 
Clearly, the normal force oscillates, but with negligible amplitude. We see that the axial strain gives tension while flexural strain results in compression force. For the $D^a$ model, these contributions have nearly equal absolute values and the normal force oscillates around zero. The result is similar for the $D^3$ model, the normal force oscillates about the value close to 0.1. For the $D^2$ model, the error becomes pronounced since the normal force oscillates around approximately 9.3. Due to the fact that the influence of flexural strain on normal section force is disregarded in the $D^1$ model, the only contribution to the normal force comes from the axial strain, which oscillates around zero for this model. This result is considered accurate, but it stems from the fact that the two errors have canceled each other out. I.e. the dilatation is erroneously obtained as near zero, and the influence of the flexural strain is disregarded. These two approximations gave the correct value of normal force. Finally, the $D^0$ model gives a completely inaccurate result. For this model, the axial strain is negative and there is no contribution from the flexural strain.

\subsubsection{Test of the algorithms for the update of internal forces}

The simplicity of this example is ultimately utilized in testing the algorithms for the update of internal stress. Three approaches are considered: (i) the current stress is calculated from the total strain, (ii) the current stress is calculated by \eqqref{eq:linearization of stress} with the first order approximation of adjustment term, and (iii) the current stress is calculated by \eqqref{eq:linearization of stress} with the zeroth order approximation of adjustment term. These approaches are designated as $F1$, $F2$ and $F3$, respectively, and they are elaborated in Appendix A.

\begin{figure}
	\includegraphics[width=\linewidth]{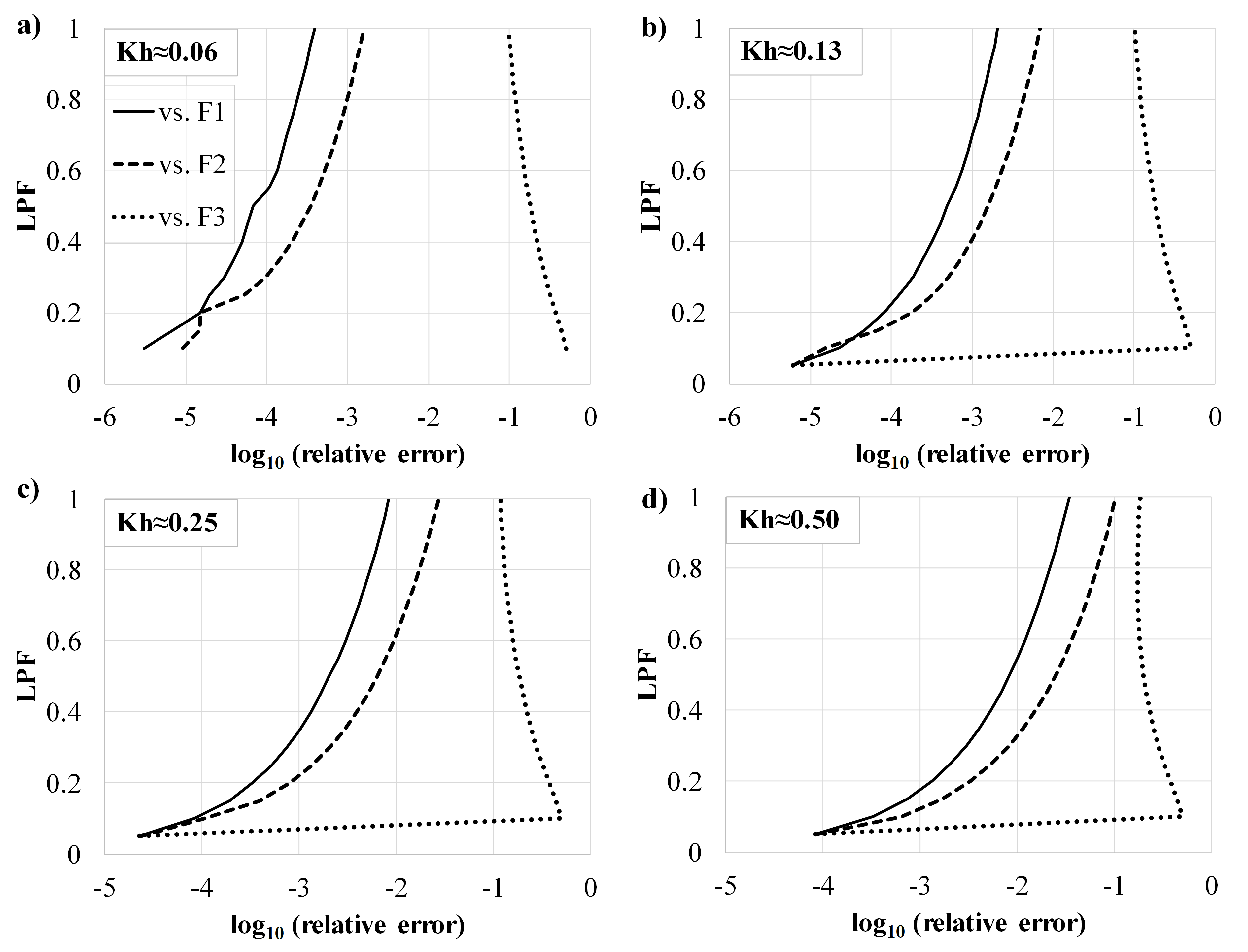}
	\caption{Cantilever beam. Comparison of relative errors of axial strain of beam axis for three algorithms for the update of internal forces. Four values of curviness are considered: a) $Kh \approx 0.06$, $Kh \approx 0.13$, $Kh \approx 0.25$, $Kh \approx 0.50$. }
	\label{fig:mainspring: mainsptring strain}
\end{figure}

The physical normal force, derived in \cite{2018borkovicb}, is:
\begin{equation}
\label{eq:n force phis}
N=E\ii{A}{}{0} \ii{\epsilon}{}{(11)} - \frac{1}{2} E \ieq{I}{}{} \chi,
\end{equation} 
where $\chi$ is the change of curvature with respect to the Frenet-Serret frame of reference \cite{2018borkovicb, 2020radenkovicb}. From the condition of pure bending ($N=0$), we can calculate the exact physical axial strain of beam axis as:
\begin{equation}
\label{eq:n exact strain og beam axis}
\ii{\epsilon}{}{(11)} = \frac{\ieq{I}{}{} \chi}{2A_0},
\end{equation}
since $\chi=K = LPF \left( 2 n \pi \slash L \right)$ for this example. This exact value of axial strain is compared with the results obtained with three different numerical approaches for the calculation of internal forces, Fig. \ref{fig:mainspring: mainsptring strain}. 
For the first load increment, which is 0.05 here, the internal forces are calculated in a same way for all algorithms. Interestingly, for the model with $Kh\approx0.06$, all three algorithms return the same result as \eqqref{eq:n exact strain og beam axis} for $LPF=0.05$. The $F1$ approach gives the most accurate results and this is also approach utilized in all present examples. The $F2$ algorithm can be successfully employed, however, for those beams with moderate curvature. The algorithm $F3$ is not recommended since it gives large errors, even for the small values of curviness. Inevitably, both $F1$ and $F2$ algorithms give erroneous results as the curviness and strain increase.

\clearpage
\subsection{Lee's frame}
\begin{figure}[t]
    \centering
	\includegraphics[width=0.975\linewidth]{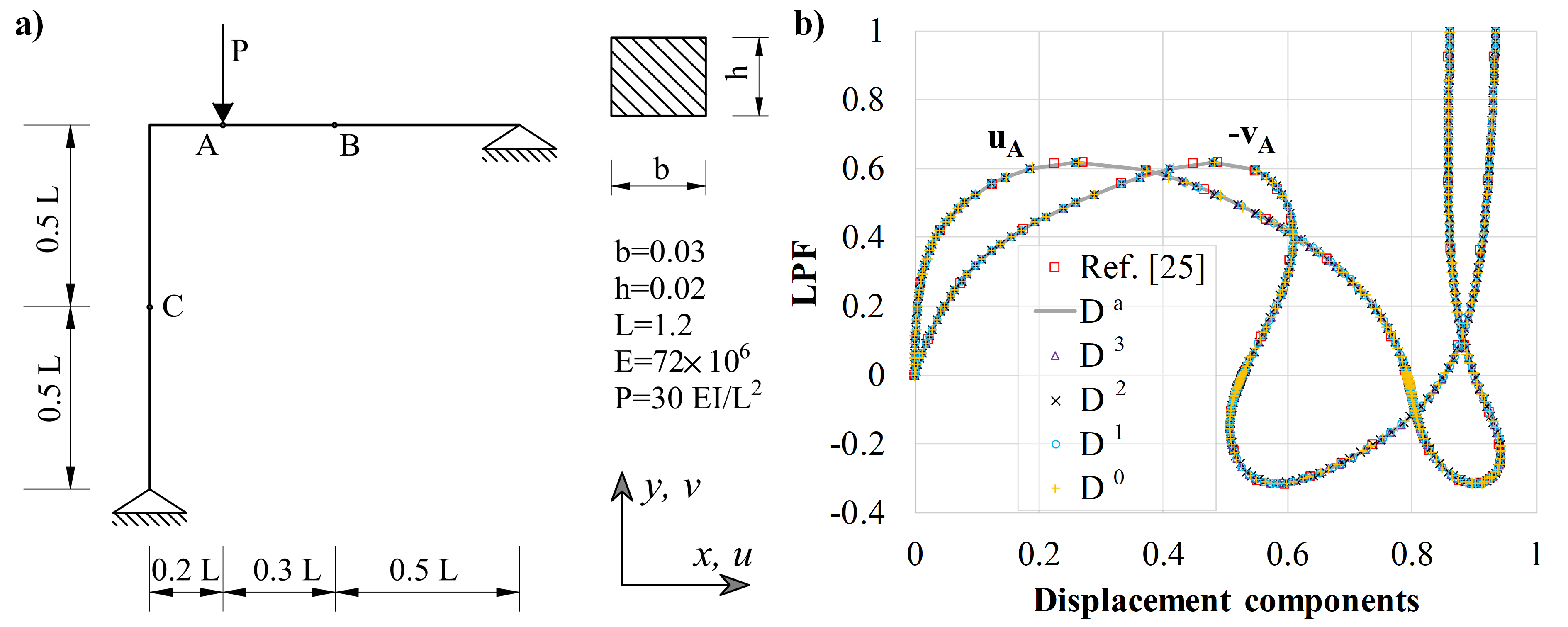}
	\caption{Lee's frame. a) Geometry and applied load. b) Comparison of the equilibrium paths for two characteristic displacement components using developed models and \cite{2020vo}. } 
	\label{fig:lee frame: disposition and eq path}
\end{figure}

\begin{figure}[b]
	\includegraphics[width=\linewidth]{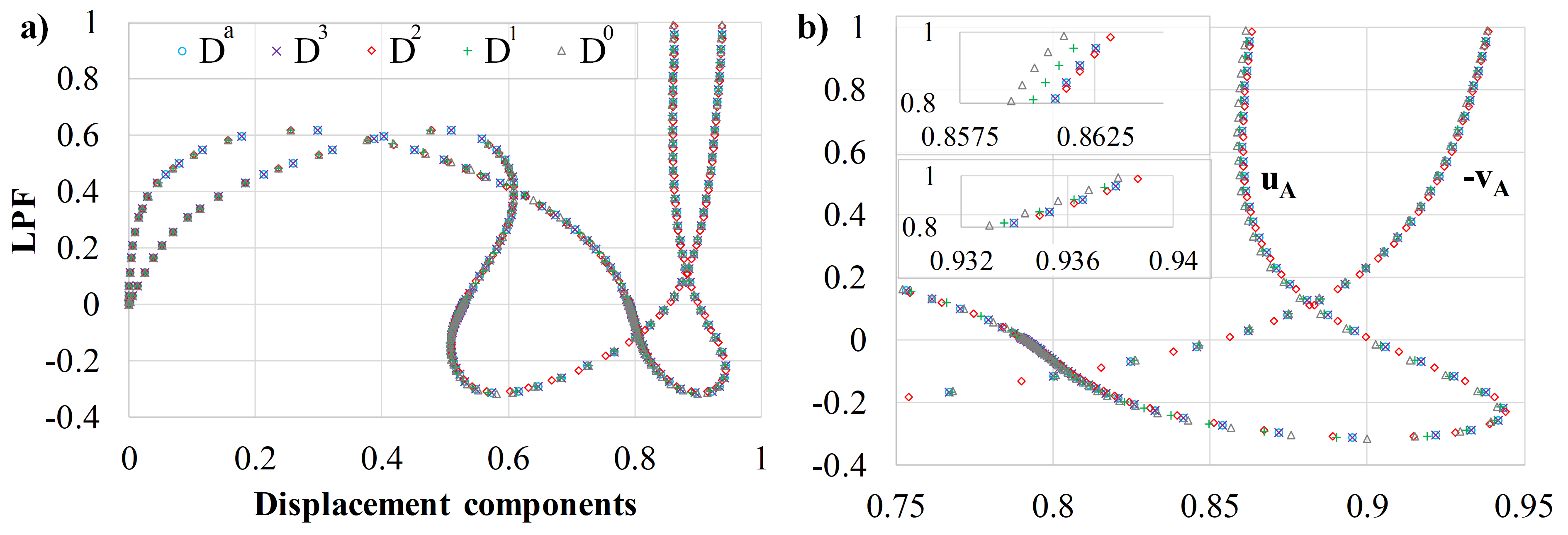}
	\caption{Lee's frame. Equilibrium paths for the frame $LF1$ using five different constitutive models: a) complete equilibrium path, b) zoomed parts of the equilibrium path.  }
	\label{fig:lee frame: h=2.5x0.02}
\end{figure}

\begin{figure}[h]
	\includegraphics[width=\linewidth]{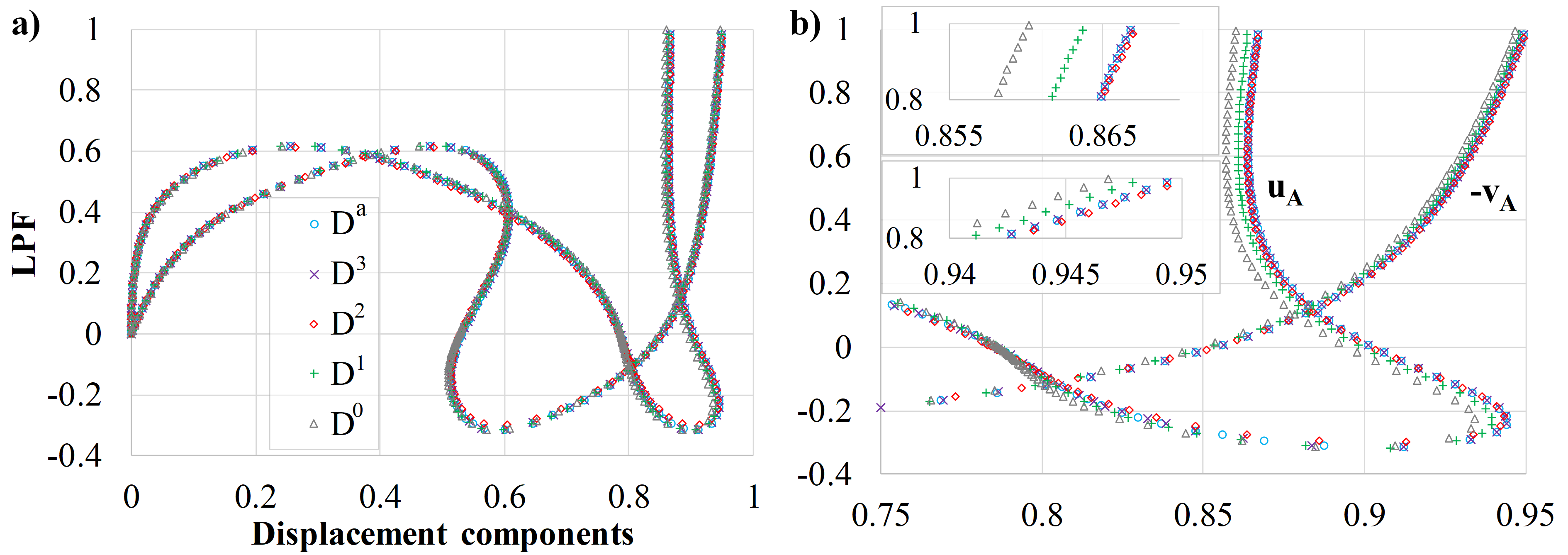}
	\caption{Lee's frame. Equilibrium paths for the frame $LF2$ using five different constitutive models: a) complete equilibrium path, b) zoomed parts of the equilibrium path.  }
	\label{fig:lee frame: h=5x0.02}
\end{figure}

\begin{figure}[h]
	\includegraphics[width=\linewidth]{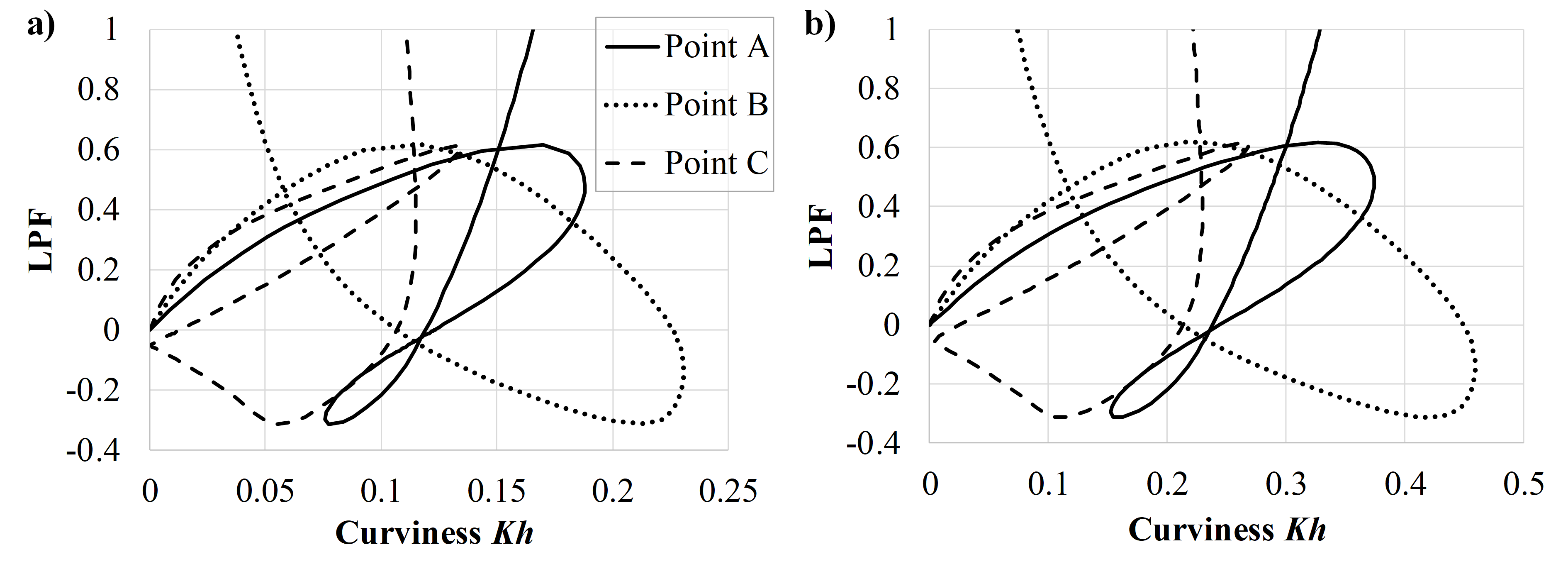}
	\caption{Lee's frame. Curviness at three characteristic sections for the frames with increased height of cross section: a) case $LF1$, b) case $LF2$. }
	\label{fig:lee frame: curviness}
\end{figure}

This example is also well-established in the literature in the context of nonlinear analysis of in-plane beams \cite{1986schweizerhof, 2016huang, 2018rezaiee-pajand}. The frame consists of two rigidly connected beams and the displacement components at the point of force application are observed, Fig.~\ref{fig:lee frame: disposition and eq path}a. Each beam is discretized with five quartic $C^1$ elements. 
The results obtained with the presented models are given in Fig.~\ref{fig:lee frame: disposition and eq path}b and they are in full agreement with those from literature \cite{2020vo}. This is expected because the maximum curviness in this example is local and less than 0.1.
In order to examine the influence of curviness on this structure, the height of the cross section is multiplied by 2.5 ($h_1=0.05$) and 5 ($h_2=0.1$), and the obtained examples are designated as $LF1$ and $LF2$, respectively. The example with the original height, $h=0.02$, is labeled $LF$. The equilibrium paths for the examples with increased height are shown in Figs.~\ref{fig:lee frame: h=2.5x0.02} and \ref{fig:lee frame: h=5x0.02}, while the curviness at three characteristic sections is given in Fig.~\ref{fig:lee frame: curviness}.
It is evident that the curviness at these sections follows similar path for both observed examples. The difference in the value of curviness practically equals the difference in cross section heights. The maximum curviness for $LF1$ is $Kh \approx 0.23$, while for $LF2$ it is $Kh \approx 0.46$.
In the context of the structural response of these frames, all constitutive relations return similar results in the case of $LF1$. inspecting the zoomed equilibrium path for the component $v_A$ allows us to estimate the difference between $D^a$ and $D^0$ models which is close to $0.2 \% $ for $LPF=1$, see Fig.~\ref{fig:lee frame: h=2.5x0.02}b. For the $LF2$ example, the discrepancies are more noticeable. The largest difference between $D^0$ and $D^a$ models is near $0.8 \% $ at the final configuration while the results returned by the $D^2$, $D^3$, and $D^a$ models are virtually indistinguishable.

\begin{figure}[h]
	\includegraphics[width=1.\linewidth]{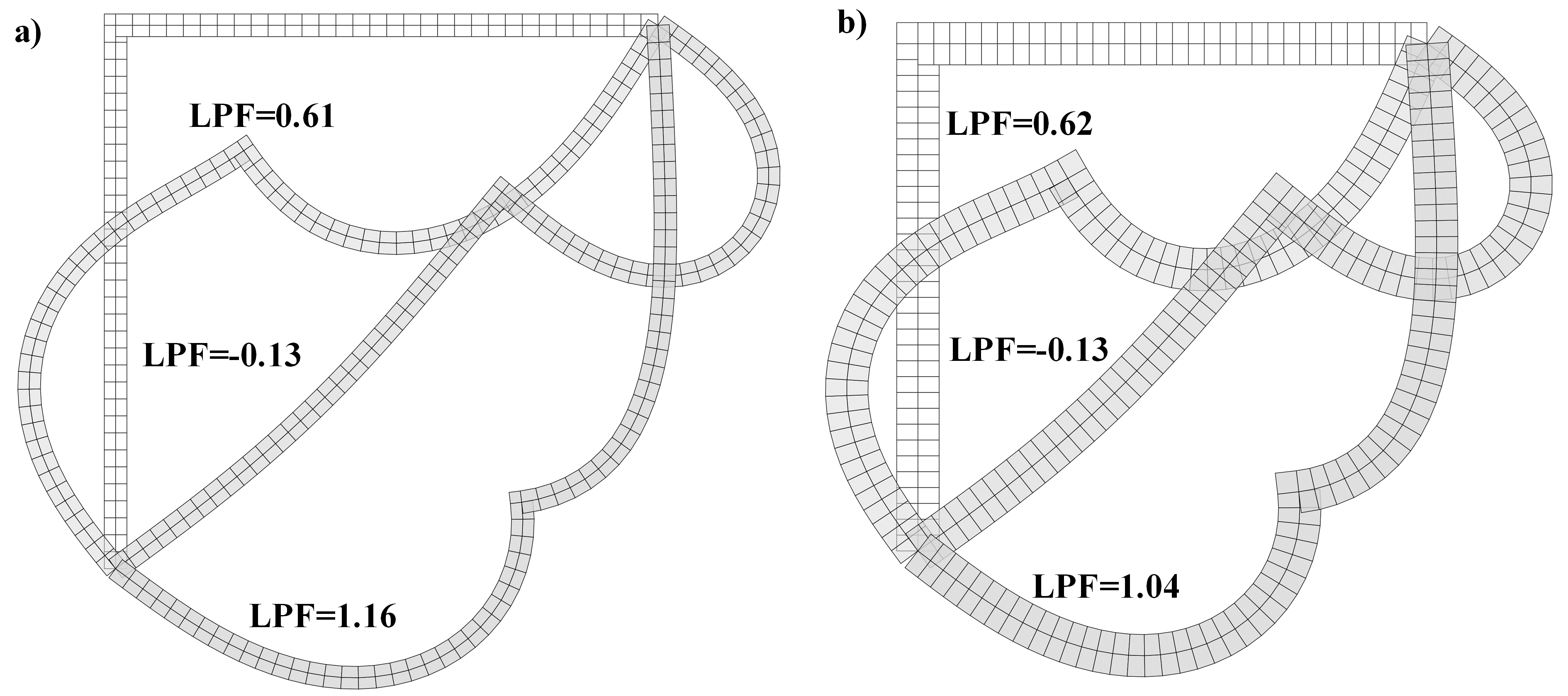}
	\caption{Lee's frame. Deformed configurations of structures with increased height for three values of $LPF$: a) case $LF1$, b) case $LF2$. }
	\label{fig:lee frame: def config}
\end{figure}

Although the maximum curviness for the examples with increased cross section height is large, the difference between displacements obtained by different constitutive models is significantly lower in comparison with the previous example, Fig.~\ref{fig:mainspring: displacement relative difference}. This is mainly due to the fact that the maximum curviness of Lee's frame is local, while in the previous example it was constant along the whole beam and, therefore, more impactful on the structural response. This local character of the maximum curviness is underlined in Fig.~\ref{fig:lee frame: def config} where the deformed configurations for $LF1$ and $LF2$ are displayed for three $LPF\textnormal{s}$.
Moreover, the rotations of adjacent sections at the joint of patches match, which confirms the correct application of the constraint equation.

\begin{figure}[h]
	\includegraphics[width=\linewidth]{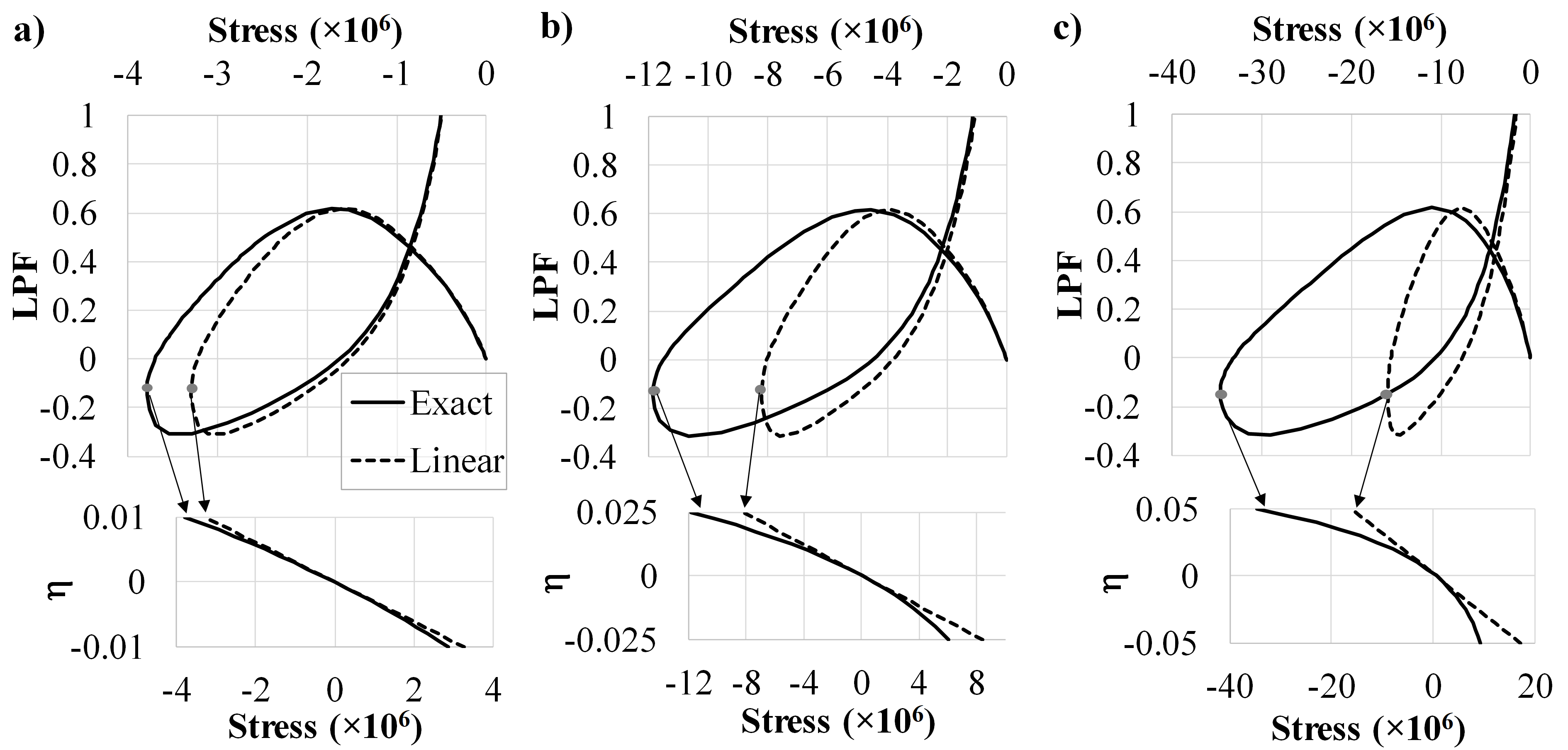}
	\caption{Lee's frame. Stress at the outer fiber at the section $B$ vs. $LPF$ (above), and the distribution of stress across the cross-section height (bellow) for designated points on equilibrium path: a) case $LF$, b) case $LF1$, c) case $LF2$.  }
	\label{fig:lee frame: stress}
\end{figure}

Finally, the stress at the section $B$ is considered in Fig.~\ref{fig:lee frame: stress}. 
The constitutive model $D^a$ is utilized for the calculation of displacements and reference strains. Then, two expressions for the distribution of stress are utilized: (i) the exact one, based on Eqs.~\eqref{eq:final form of eq strain} and \eqref{eq:stress strain relation}, and (ii) the linear one, based on the same equations but with $K=0$. These distributions are shown below the equilibrium paths for the stress at outer fiber in Fig.~\ref{fig:lee frame: stress}. It is clear that, the exact stress distribution should be utilized in a post-processing phase even for structures with a maximum local curviness less than 0.1, such as the original example $LF$. The difference of extreme stresses between the exact and the linear stress distributions is close to $13 \%$ for the $LF$ example. As the curviness increases, this error becomes more pronounced.

\clearpage
\subsection{Multi-snap behavior of a parabolic arch}

\begin{figure}
	\includegraphics[width=\linewidth]{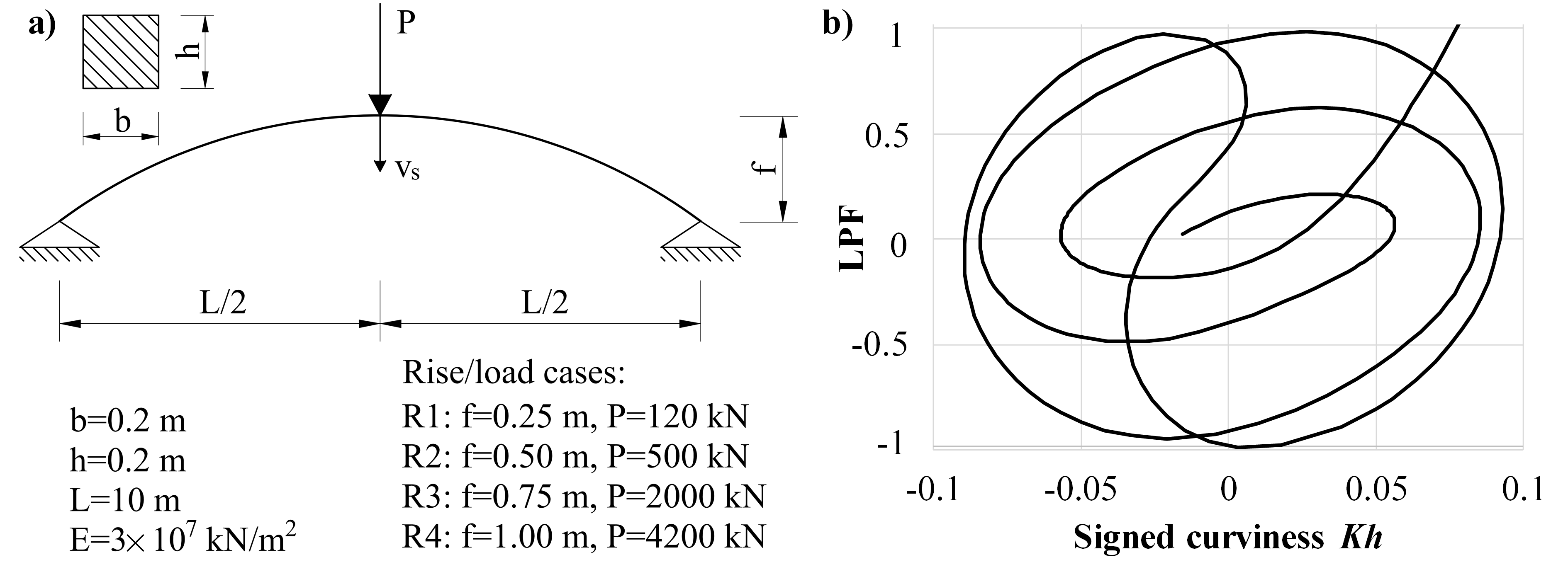}
	\caption{Parabolic arch. a) Geometry and applied load; b) Signed curviness vs. $LPF$ for $R4$ case. }
	\label{fig:multi1}
\end{figure}

The final example deals with the nonlinear response of a parabolic arch depicted in Fig.~\ref{fig:multi1}a. 
The arch is simply supported and loaded with the vertical force at the apex. For shallow arches, the snap-through phenomenon occurs after which the structure reaches a stable equilibrium and shows stiffening behavior. However, for deeper arches, a multiple snap behavior can be expected \cite{1973sabir, 1978harrison, 1990clarke, 1990yang, 2017radenkovic}. The aim of this example is to show that the present formulation is capable of describing highly-complex responses of arbitrarily curved structures. Furthermore, we discuss a physical reasoning behind the observed behavior \cite{1973sabir}. We consider four parabolic arches labeled $R1$, $R2$, $R3$, and $R4$. They differ in rise $f$ and the magnitude of the applied load $P$, as detailed in Fig.~\ref{fig:multi1}a. Due to the symmetry, one half of the arch is modeled with 16 cubic $C^2$ elements. As a guiding solution, we use an Abaqus simulation, which employs about 30 straight cubic B23 elements \cite{2009smith}. It is worth noting that Abaqus could not converge with denser meshes. 

Fig.~\ref{fig:multi1}b shows the signed curviness at the apex for the beam with the largest curvature, i.e., R4. The graph is interesting due to the multiple limit points. The important fact to note is that the maximum local curviness of this beam is lower than 0.1. 

\begin{figure}[b!]
\includegraphics[width=\linewidth]{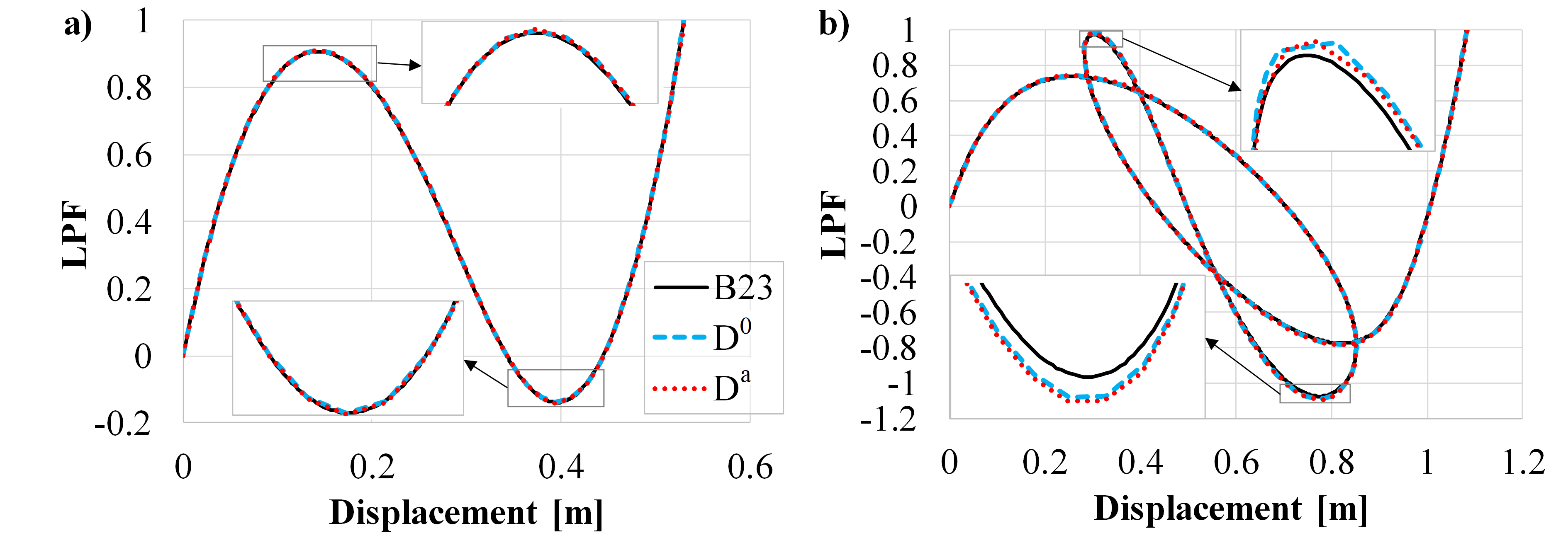}
\caption{Parabolic arch. Equilibrium paths for: a) $R1$ case, and b) $R2$ case. }
\label{fig:multi2}
\end{figure}

\begin{figure}
	\includegraphics[width=\linewidth]{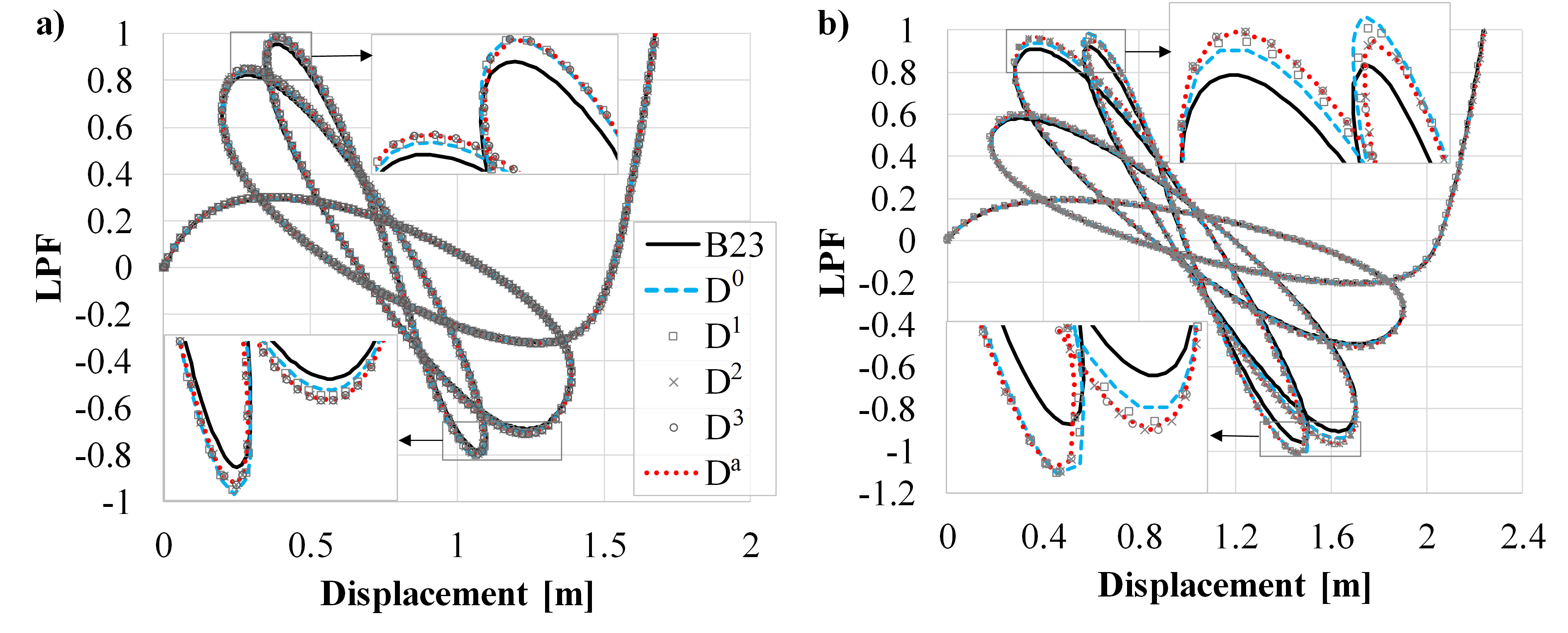}
	\caption{Parabolic arch. Equilibrium paths for: a) $R3$ case, and b) $R4$ case. }
	\label{fig:multi3}
\end{figure}

For the $R1$ case, a relatively simple structural response is obtained, and the equilibrium path for $v_s$ is shown in Fig.~\ref{fig:multi2}a. 
As the rise increases to 0.5 m, one snap-back occurs, Fig.~\ref{fig:multi2}b. Since these two arches have curviness less than 0.025, only the results obtained with the simplest, $D^0$, and the most elaborated, $D^a$, models are displayed. They agree well for almost the complete equilibrium path while negligible discrepancies can be observed at the load limit points. Also, the results for the $R1$ case are practically indistinguishable from those obtained by Abaqus. Regarding the $R2$ case, small differences occur at the load limit points. 
As the rise increases further, the arch behavior becomes more complex and multiple snaps occur, see Fig.~\ref{fig:multi3}. 
Since the results for the different constitutive models are in good agreement, it can be concluded that the influence of the curviness is not significant for these arches. However, an inspection of the equilibrium paths near the load limit points reveals that a difference between the present approach and the B23 element discretization exists, which increases with the curviness and complexity of the equilibrium path. The results obtained by the $D^3$ and $D^2$ models are practically identical to those of the $D^a$ model. Furthermore, the models $D^0$ and $D^1$ deviate slightly from these results in the vicinity of limit points, especially for the $R4$ case. It is noteworthy that all models return the same final deformed configuration.

\begin{figure}
	\includegraphics[width=\linewidth]{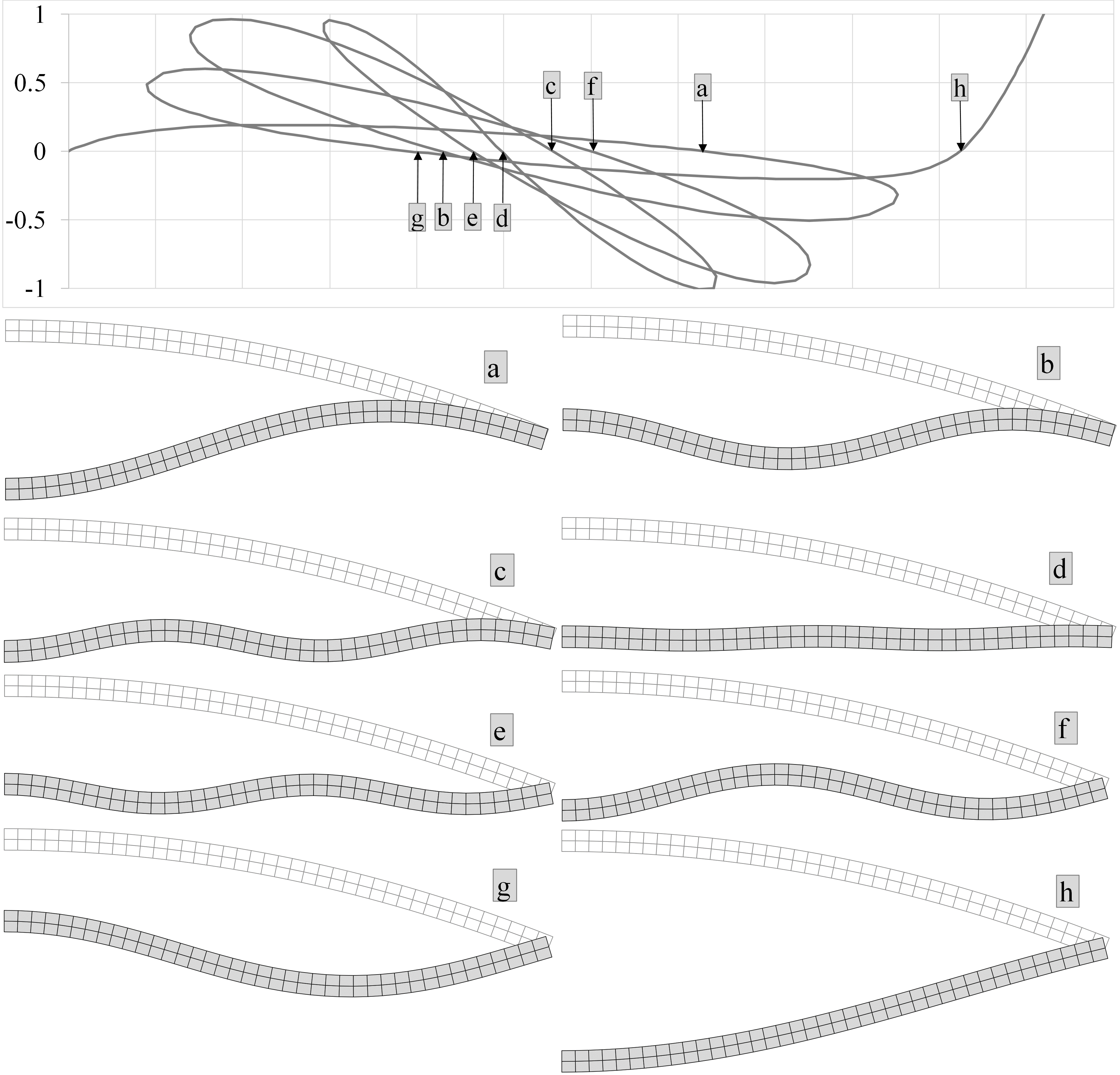}
	\caption{Parabolic arch. Deformed configurations of $R4$ case for $LPF=0$.  }
	\label{fig:multi4}
\end{figure}

In order to analyze this multi-snap phenomenon more thoroughly, the eight different deformed configurations of the $R4$ case at $LPF=0$ are given in Fig.~\ref{fig:multi4}. 
It can be seen that these configurations are similar to the buckling/vibration eigenmodes of a simple straight beam. In particular, the configurations designated with $a$, $b$, $c$, and $d$ resemble the $3^{rd}$, $5^{th}$, $7^{th}$, and $9^{th}$ beam eigenshape, respectively. Each of these configurations is stiffer than the previous one and the absolute value of load limit points increases as well. After configuration $d$, which is seemingly straight, the arch cannot generate any more half-waves and starts to release accumulated strain energy. Hence, the point $d$ in Fig.~\ref{fig:multi4} approximately represents an inflection point of the equilibrium path. After this the load limit points decrease and the arch passes through all the previous configurations in reverse order. Finally, a stable equilibrium branch is reached after the last load limit point.

It should be noted that this multi-snap behavior depends heavily on the imposed conditions of symmetry. In reality some small perturbations would probably exist in either the force position or the geometry, which would result in a simpler response \cite{1978harrison}.

\begin{figure}
	\includegraphics[width=\linewidth]{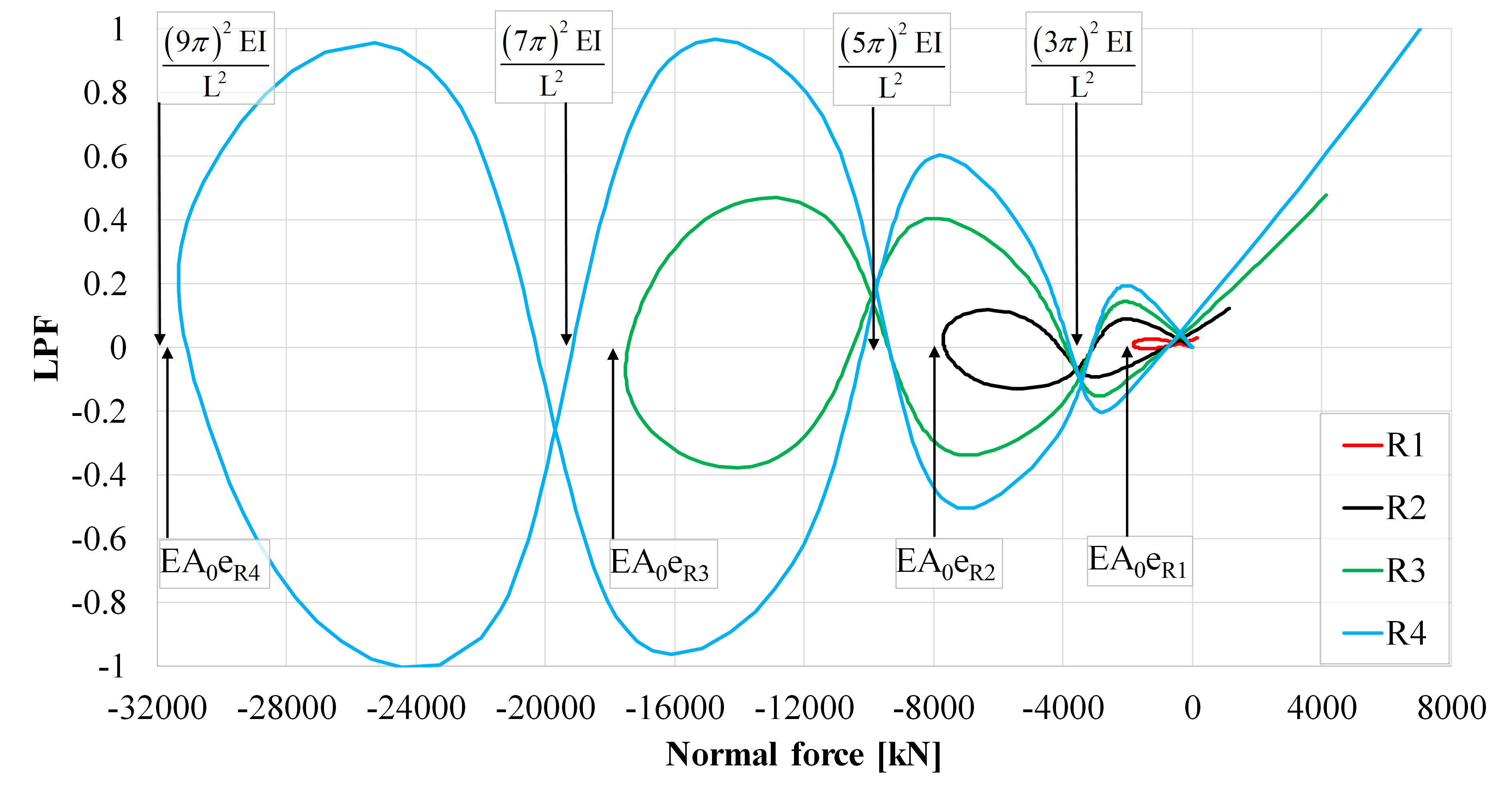}
	\caption{Parabolic arch. Normal force at the apex for all considered cases. Critical buckling forces and estimated maximum normal compression forces $\left( \textnormal{EA}_0 \textnormal{e}_{\textnormal{Rn}} \right)$ are marked by arrows.}
	\label{fig:multi5}
\end{figure}

The detected behavior is further investigated by observing the normal force at the apex. The applied forces for all cases are scaled in relation to the largest one. Fig.~\ref{fig:multi5} illustrates the resulting equilibrium paths and marks the values of $3^{rd}$, $5^{th}$, $7^{th}$, and $9^{th}$ critical buckling loads of a simply supported, axially loaded beam with length $L$.
Additionally, an estimate of the maximum possible axial strain of the beam axis is taken into consideration by calculating the initial length of the parabolic arch $\idef{L}{}{}$ and by finding the Green-Lagrange strain:
\begin{equation}
\ii{e}{}{Rn}=\frac{\idef{L}{2}{n}-10^2}{2 \cdot 10^2}.
\end{equation}
This is the strain of the parabola which deforms into a straight line. Multiplication of this strain with the axial stiffness $EA_0$ gives an estimate of a maximum normal compression force that can be generated in these arches due to the considered applied load. These forces are also designated in Fig. \ref{fig:multi5}. 

The number of half-wavelength forms that an arch can attain during this symmetric snapping is limited. This limit is directly influenced by the length of the arch and the critical buckling forces of a simple straight beam with length $L$. To summarize, the arch $R4$ cannot make the $11^{th}$ eigenshape, since its maximum axial strain cannot generate the normal force that is large enough to snap the beam into this mode. The analogous conclusion is valid for all observed cases. When there are no more new stable forms to reach, the curvature of equilibrium path changes sign and the arch goes through all these configurations in reverse, until it finds the stable form. A similar discussion takes place in \cite{1973sabir}, where the rise, thrust and flexural stiffness are identified as the causes for multi-snap behavior of circular arches. However, the graphical description given in Fig. \ref{fig:multi5} provides additional insight into the nature of the phenomenon.

\section{Conclusions}

A rigorous metric of the plane BE beam is utilized consistently for the derivation of the weak form of the equilibrium. The spatial discretization of the virtual power is performed by IGA. The introduction of the full beam metric provides a higher-order accurate BE beam formulation. Four simplified models are derived in addition to the exact constitutive relation. The comparison of these models via numerical examples allows a detailed analysis of the influence the formulation components have and of the most importance, the beam's curviness. The present formulation is geometrically exact, rotation-free, and fully capable of dealing with complex responses of multi-patch structures.

The results show that, in order to correctly determine the axial strain at the centroid of a curved beam, a rigorous computational model must be employed. For beams with curviness $Kh<0.1$, the simple decoupled equations return reasonably accurate results for the displacement field. As the curviness increases, its influence becomes noticeable and as a result a more involved and complex model is required. An important factor is the domain where the strong curviness exists. If this is relatively local, its influence on the global response will not be as significant as when large portions of the structure are strongly curved. Regardless of the constitutive model, utilizing the exact expressions for the equidistant strains and stresses is recommended, since these have a nonlinear distribution and this even for beams with a small curvature. The inclusion of these expressions in a post-processing phase is a simple method for improving accuracy.

A phenomenon of multi-snap behavior of a parabolic arch is revised and elaborated. The arch accumulates strain energy and deforms into the configurations that resemble the eigenmodes of a straight beam. After the limit eigenshape is reached, the arch releases the accumulated strain energy and finds a stable equilibrium branch.

An interesting direction for future research is the application of the proposed formulation to the statics and dynamics of spatial beams.

\section*{Acknowledgments}

During this work, our beloved colleague and friend, Professor Gligor Radenkovi\'{c} (1956-2019), passed away. The first author acknowledges that his unprecedented enthusiasm and love for mechanics were crucial for much of his previous, present, and future research.

We acknowledge the support of the Austrian Science Fund (FWF): M 2806-N.

\section*{Appendix A. Update of internal forces - fix formula}
\setcounter{equation}{0}
\renewcommand\theequation{A\arabic{equation}}

Due to the effect of curviness, the update of internal forces requires special attention. In order to add the increment of stress as in \eqqref{eq:linearization of stress}, the cumulative stress from the previous configuration must be adjusted to the metric of the current configuration, as discussed in Section 4.1. After the addition, the stress is integrated and the stress resultant and stress couple are obtained. First, let us find the relation between previously calculated stress with respect to two configurations. By assuming that the unit tangent vector of the beam axis does not change its direction significantly between configurations $(n)$ and $(n+1)$, we obtain:
\begin{equation}
\label{eq:traction}
\iiieq{\bm{t}}{(n)}{}{}{} = \left( \iiieq{\sigma}{(n)}{(n)}{11}{} \: \iiieq{\bm{g}}{(n)}{}{}{1} \otimes \iiieq{\bm{g}}{(n)}{}{}{1} \right) \frac{\iiieq{\bm{g}}{(n)}{}{}{1}}{\sqrt{\iiieq{g}{(n)}{}{}{}}}  \approx \left( \iiieq{\sigma}{(n)}{(n+1)}{11}{} \: \iiieq{\bm{g}}{(n+1)}{}{}{1}
\otimes \iiieq{\bm{g}}{(n+1)}{}{}{1} \right) \frac{\iiieq{\bm{g}}{(n+1)}{}{}{1}}{\sqrt{\iiieq{g}{(n+1)}{}{}{}}}.
\end{equation}
This assumption allows us to write the required relation as:
\begin{equation}
\label{eq:traction1}
\iiieq{\sigma}{(n)}{(n)}{11}{} \: \iiieq{g}{(n)}{}{}{} \approx \iiieq{\sigma}{(n)}{(n+1)}{11}{} \: \iiieq{g}{(n+1)}{}{}{}
\implies \iiieq{\sigma}{(n)}{(n+1)}{11}{} \approx \frac{(\iii{g}{(n)}{}{}{0})^2 \: \iii{g}{(n)}{}{}{}} {(\iii{g}{(n+1)}{}{}{0})^2 \: \iii{g}{(n+1)}{}{}{}} \: \iiieq{\sigma}{(n)}{(n+1)}{11}{}.
\end{equation}
The adjustment term next to $\iiieq{\sigma}{(n)}{(n+1)}{11}{}$ consists of the ratio of the initial curvature correction factors at two configurations which makes it inappropriate for exact integration in nonlinear beam analysis. Therefore, some approximation is required. The first order Taylor approximation of this ratio yields:
\begin{equation}
\label{eq:traction2}
\left( \frac{\iii{g}{(n)}{}{}{0} } {\iii{g}{(n+1)}{}{}{0} } \right) ^2 =\left( \frac{1-\eta \iii{K}{(n)}{}{}{}}{1-\eta \left( \iii{K}{(n)}{}{}{} + \iii{\chi}{(n+1)}{}{}{} \right)} \right)^2 \approx  1+ 2\eta\iii{\chi}{(n+1)}{}{}{},
\end{equation}
where $\iii{\chi}{(n+1)}{}{}{}$ is the change of curvature in the current increment with respect to the Frenet-Serret frame of reference. With this approximated adjustment term, we can perform the integration, but the higher order terms with respect to $\eta$ must be disregarded. By this means the energetically conjugated section forces from the previous configuration are adjusted to the metric of the current configuration using Eqs.~\eqref{eq: rates of section forces}, \eqref{eq:traction1}, and \eqref{eq:traction2}:
\begin{equation}
\label{eq: update of sf}
\begin{aligned}
\iiieqt{N}{(n)}{(n+1)}{}{} &\approx \left(1+ 2\eta\chi \right) \iiieqt{N}{(n)}{(n)}{}{} \approx \iiieqt{N}{(n)}{(n)}{}{} - 2 \chi \iiieqt{M}{(n)}{(n)}{}{}, \\
\iiieqt{M}{(n)}{(n+1)}{}{} &\approx \iiieqt{M}{(n)}{(n)}{}{}.
\end{aligned}
\end{equation}
Evidently, the energetically conjugate bending moment does not require adjustment, while the normal force does.

This approach for the update of internal forces is the rational one, since the integration across the cross section should be avoided at each increment. An alternative method, that proved even more accurate in this research, is to calculate the internal forces from the total strain and the current constitutive matrix. However, this approach is not theoretically sound since the linear constitutive relation is valid only for the increments of strain and stress. Nevertheless, if the strains are not large, the results in Section 5 suggest that this approach can return acceptably accurate results. As noted previously, the latter approach is designated as $F1$ and former one as $F2$. Additionally, an incremental approach for the update of internal forces without any adjustment of previously calculated forces is implemented and marked as $F3$. When observing the \eqqref{eq:traction2}, it can in fact be considered as the zeroth order approximation. The results of these three approaches are compared in Subsection 5.1.4.

\bibliography{beam} 
\bibliographystyle{ieeetr}

\end{document}